\documentclass{article}

\usepackage{microtype}
\usepackage{graphicx}
\usepackage{booktabs}
\usepackage[hyphens]{url}
\usepackage{subcaption} 
\usepackage{standalone}
\usepackage{float}
\usepackage{xspace}
\usepackage{fancyvrb}
\usepackage{algorithm} 
\usepackage{amsmath}
\usepackage{pgf}
\usepackage{import}
\usepackage{lmodern} 
\usepackage{multirow}
\usepackage[most]{tcolorbox}
\usepackage{listings}
\usepackage[table]{xcolor}
\usepackage{enumitem}
\usepackage{subcaption}
\usepackage{graphicx}
\usepackage{makecell}
\usepackage{adjustbox}
\usepackage{caption}
\usepackage{hyperref}
\usepackage{chngcntr}

\usepackage{colortbl}

\definecolor{headercolor}{RGB}{220,230,241}
\definecolor{highcolor}{RGB}{235,245,255}
\definecolor{lowcolor}{RGB}{235,255,235}
\definecolor{midcolor}{RGB}{255,245,235}

\usepackage{eso-pic}
\usepackage{xparse}
\newlength{\badgewidth}
\newlength{\badgegap}
\newcommand{\badgeList}{}

\NewDocumentCommand{\addTopRightBadge}{m}{%
  \gappto{\badgeList}{%
    \includegraphics[width=\badgewidth,keepaspectratio]{#1}%
    \hspace{\badgegap}%
  }%
}

\newcommand{\placeTopRightBadges}[1]{%
\AddToShipoutPictureBG*{%
\put(\LenToUnit{\paperwidth - 0.5cm - \badgewidth},%
     \LenToUnit{\paperheight - 2.5cm}){%
\makebox[0pt][r]{\href{#1}{\badgeList}}%
}%
}%
}
\addTopRightBadge{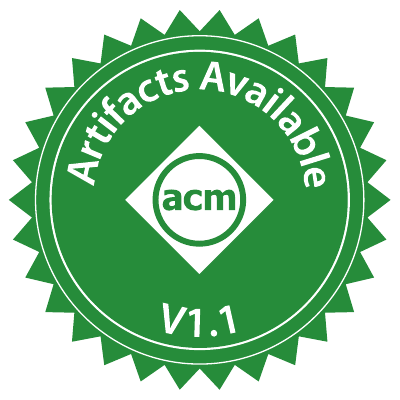}
\addTopRightBadge{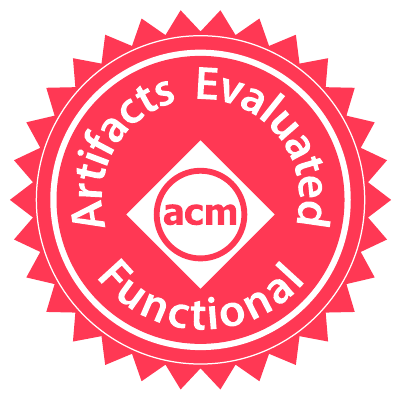}
\addTopRightBadge{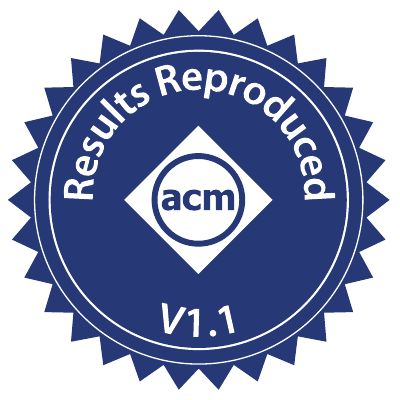}

\placeTopRightBadges{https://www.acm.org/publications/policies/artifact-review-and-badging-current}

\usepackage[accepted]{mlsys2026}

\definecolor{codegreen}{rgb}{0,0.6,0}
\definecolor{codegray}{rgb}{0.5,0.5,0.5}
\definecolor{codepurple}{rgb}{0.58,0,0.82}
\definecolor{backcolour}{rgb}{0.95,0.95,0.92}
\definecolor{bg}{RGB}{250,250,250}
\setlist[itemize]{topsep=1pt,itemsep=1pt,parsep=0pt,partopsep=0pt}
\setlist[enumerate]{topsep=1pt,itemsep=1pt,parsep=0pt,partopsep=0pt}

\newif\ifmainpaper
\newif\ifappendixonly
\newif\iffullpaper

\fullpapertrue

\usepackage{minted}
\setminted{
        linenos=true,
        autogobble,
        breaklines=true,    
}

\newenvironment{code}
{\minted[escapeinside=@@,xleftmargin=15pt,fontsize=\scriptsize,breakanywhere=true,baselinestretch=0.9]{cuda}}
{\endminted}
\lstdefinelanguage{CUDA}{
    language=C++,
    morekeywords={
        __global__, __device__, __shared__, __host__, __constant__,
        __syncthreads, threadIdx, blockIdx, blockDim, gridDim,
        int4, float4, dim3
    },
    keywordstyle=\color{red!60!black}\bfseries,
    commentstyle=\color{green!40!black}\itshape,
    backgroundcolor=\color{bg},   
    numberstyle=\tiny\color{codegray},
    stringstyle=\color{codepurple},
    basicstyle=\ttfamily\footnotesize,
}

\tcbset{
    kernelbox/.style={
        enhanced,
        colback=black!2,
        colframe=black!40,
        boxrule=0.4pt,
        arc=2pt,
        left=3pt,
        right=3pt,
        top=3pt,
        bottom=3pt,
        listing only,
        width=\columnwidth,
        breakable
    }
}

\newtcblisting{cudakernel}[2][]{
    kernelbox,
    title={#2},
    listing options={
        language=CUDA,
        basicstyle=\ttfamily\scriptsize,
        numbers=left,
        numberstyle=\tiny\color{gray},
        breaklines=true,
        breakatwhitespace=false,
        columns=fullflexible,
        keepspaces=true,
        showstringspaces=false,
        tabsize=2,
        literate=
            {*}{{\char42}}1
            {+}{{\char43}}1
            {-}{{\char45}}1
            {/}{{\char47}}1
    },
    #1
}

\lstdefinestyle{bash}{
    language=bash,
    morekeywords={sudo, apt-get, git, make, conda, pip3, bash, cd, install},
    basicstyle=\rmfamily\footnotesize,
    backgroundcolor=\color{bg},
    keywordstyle=\color{red!60!black},
    commentstyle=\color{green!40!black}\itshape,
    stringstyle=\color{blue!70!black},
    showstringspaces=false,
    breaklines=true,
    frame=single,
    rulecolor=\color{black},
    columns=fullflexible,
    keepspaces=true,
    morecomment=[l]{\#},
    aboveskip=4pt,
    belowskip=0pt,
}

\usepackage{newfloat}

\DeclareFloatingEnvironment[
  fileext=lop,
  listname={List of Listings},
  name=Listing
]{listingfloat}

\usepackage{threeparttable}
\usepackage{arydshln}
\definecolor{headerblue}{HTML}{EAF1FB}
\definecolor{rowalt}{HTML}{F7F9FC}
\definecolor{bestgreen}{HTML}{DDF3E4}
\definecolor{besttext}{HTML}{1F6B3B}

\mlsystitlerunning{\textrm{TokenWeave}: Efficient Compute-Communication Overlap for Distributed LLM Inference}
\newcommand{\sysname}{\textrm{TokenWeave}\xspace}
\newcommand{\Tokensplitting}{\textit{Token-Splitting}\xspace}

\newcommand{\Smartsplitting}{\textit{Smart-splitting}\xspace}
\newcommand{\smartsplitting}{\textit{smart-splitting}\xspace}

\newcommand{\tokensinbatch}{tokens-in-batch\xspace}

\makeatletter
\renewcommand{\sectionautorefname}{\S\@gobble}
\renewcommand{\subsectionautorefname}{\S\@gobble}
\renewcommand{\subsubsectionautorefname}{\S\@gobble}
\makeatother

\begin{document}

\twocolumn[
\mlsystitle{\textrm{TokenWeave}: Efficient Compute-Communication Overlap for Distributed LLM Inference}

\begin{mlsysauthorlist}
    \mlsysauthor{Raja Gond}{msr}
    \mlsysauthor{Nipun Kwatra}{msr}
    \mlsysauthor{Ramachandran Ramjee}{msr}
    \end{mlsysauthorlist}
    
    \mlsysaffiliation{msr}{Microsoft Research India}
    \mlsyscorrespondingauthor{Raja Gond}{raja.gond@outlook.com}
    \mlsyskeywords{Machine Learning, MLSys}
    
    \vskip 0.3in
    \setcounter{footnote}{1}
    \begin{abstract}
\label{sec:abstract}
Distributed inference of large language models (LLMs) using tensor parallelism can introduce communication overheads of $20$\% even over GPUs connected via NVLink, a high-speed GPU interconnect. Several techniques have been proposed to mitigate these overheads by decomposing computations into smaller tasks and overlapping communication with these subtasks. However, none of these techniques are turned on by default during tensor-parallel serving in systems like vLLM, SGLang and TensorRT-LLM. This is because the number of tokens processed per iteration is typically kept small to support low-latency serving, and decomposing such smaller workloads to enable communication overlap results in worse performance. Further, the communication itself uses many streaming multiprocessors (SMs) that would otherwise be available for computation, increasing overhead.

We present \sysname, the first system to enable efficient compute-communication overlap for tensor-parallel model inference for token lengths as small as 1024. \sysname identifies RMSNorm, a previously overlooked operation, as crucial and optimizes it along with communication by implementing a novel fused \textbf{AllReduce--RMSNorm}\footnotemark{} kernel. Further, this kernel leverages the NVSHARP/Multimem feature available on modern GPUs (e.g., Hopper, Blackwell) to jointly perform communication and RMSNorm efficiently using only $2\text{--}8$ streaming multiprocessors (SMs) on an $8\times$H100 DGX system. Our evaluations demonstrate up to $\boldsymbol{1.28\times}$ speedup in latency (baseline$\div$ours) and up to $\boldsymbol{1.19\times}$ higher throughput (ours$\div$baseline) across multiple models and workloads. In several settings, \sysname delivers better performance than an equivalent model with all communication removed. The source code is available at \url{https://github.com/microsoft/tokenweave}.
\end{abstract}
]

\printAffiliationsAndNotice{}

\section{Introduction}
\label{sec:introduction}
\begin{figure}[t]
    \centering
    \includegraphics[width=0.98\linewidth]{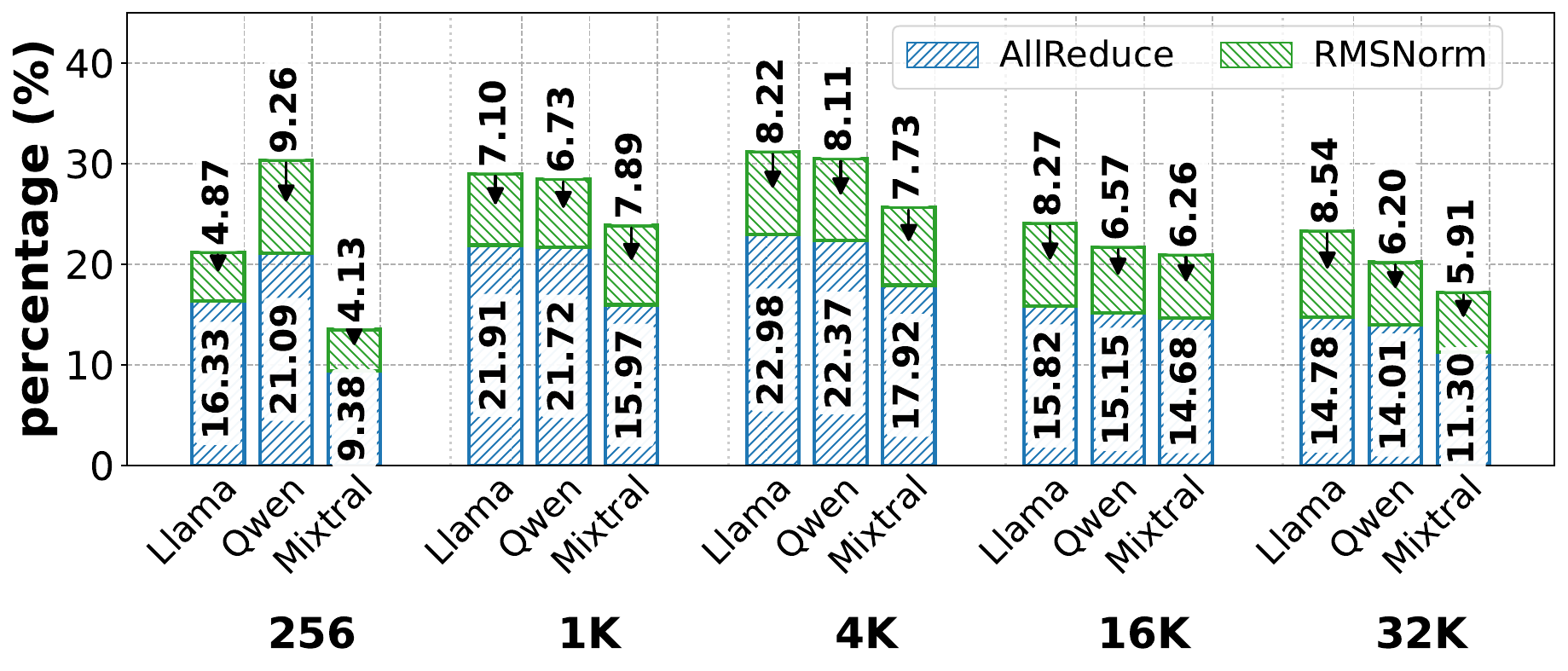}
    \vspace{-6pt}
    \caption{AllReduce communication and RMSNorm overheads for three models versus sequence length on an $8\times$H100 DGX system (median across 12 runs; GPU clocks are locked to the TDP frequency~\cite{nvidia2022setstablepowerstate}). Despite NVLink and NVSHARP, communication overheads range from $9\text{--}23$\%. RMSNorm overheads are also non-trivial, ranging from $4\text{--}9$\%.}
\vspace{-10pt}
\label{fig:allreduce-rmsnorm-cost}
\end{figure}

In recent years, large language models (LLMs) have become increasingly powerful, driven by rapid scaling of model size and training data~\cite{openai2022gpt4techreport,kaplan2020scalinglaws}. However, this trend toward larger models introduces significant challenges for efficient production deployment. Modern LLM deployments typically rely on distributed inference across multiple GPUs~\cite{megatron}, due to model size constraints or strict latency requirements. Even with high-speed interconnects like NVLink, communication overheads remain a major performance bottleneck. For example, \autoref{fig:allreduce-rmsnorm-cost} shows that communication accounts for $9\text{--}23$\% of the end-to-end inference latency for Llama-3.3-70B~\cite{llama3}, Qwen2.5-72B~\cite{qwen2.5}, and Mixtral-8x22B~\cite{mixtral} models running on an $8\times$H100 DGX using vLLM (0.8.5)~\cite{vllm} V1 engine~\cite{vllmv1}. Note that numbers are reported with GPU clocks set to their TDP frequency~\cite{nvidia2022setstablepowerstate} to ensure uniform numbers across runs in the paper unless otherwise stated. Additionally, we use instruction-tuned variants for all evaluated models.
\footnotetext{In all modern implementations, residual addition is fused with RMSNorm, and all our results use this fused implementation. For brevity, we mention only RMSNorm in all our discussions.}

To mitigate non-trivial communication overheads in distributed LLM inference, numerous solutions have been proposed in the literature. These works typically decompose computation into subtasks and overlap communication with these subtasks. Decomposition can be done either in a fine-grained manner (i.e., tile-level)~\cite{coconet, googleoverlap, flux, comet, tritondistributed, tilelink} or in a coarse-grained manner (i.e., token-level)~\cite{pipemoe, lancet, nanoflow, tbo}. However, to the best of our knowledge, none of the open-source serving systems like vLLM, SGLang~\cite{sglang} and TensorRT-LLM~\cite{tensorrtllm} enable compute-communication overlap by default for tensor-parallel inference.

This limitation arises from a {\it fundamental challenge:} communication overhead is hidden via decomposing a large computation into multiple smaller computations, but decomposition itself can suffer from a loss of compute efficiency. Modern GPUs offer a high degree of parallelism, which makes running a large computation much more efficient than running multiple smaller computations due to wave quantization effects. The decomposition overhead becomes more severe as the problem size reduces. Since AllReduce communication used in tensor-parallel deployments is highly optimized ($9\text{--}23$\% overhead, \autoref{fig:allreduce-rmsnorm-cost}), even small decomposition overheads easily overwhelm any gains from overlapping communication\footnote{Unlike tensor-parallel communication, DeepSeek's expert-parallel deployments' all-to-all has $50+$\% overhead (TBO ablation in~\cite{sglang-tbo}) and is thus easier to overlap.}. As a result, prior solutions work well only for batches with a large number of tokens ($8K+$)~\cite{tilelink}. However, such large batches are typical only in training but not in low-latency inference serving, where the number of tokens processed in each iteration is tightly constrained (e.g., \texttt{vLLM}~0.8.5 uses a default chunk size of $2048$~\cite{vllmchunkedprefill}). Thus, despite extensive prior research, tensor-parallel model inference continues to pay up to $20$\% communication costs today.

\begin{figure}[tb]
    \centering
    \includegraphics[width=\linewidth]{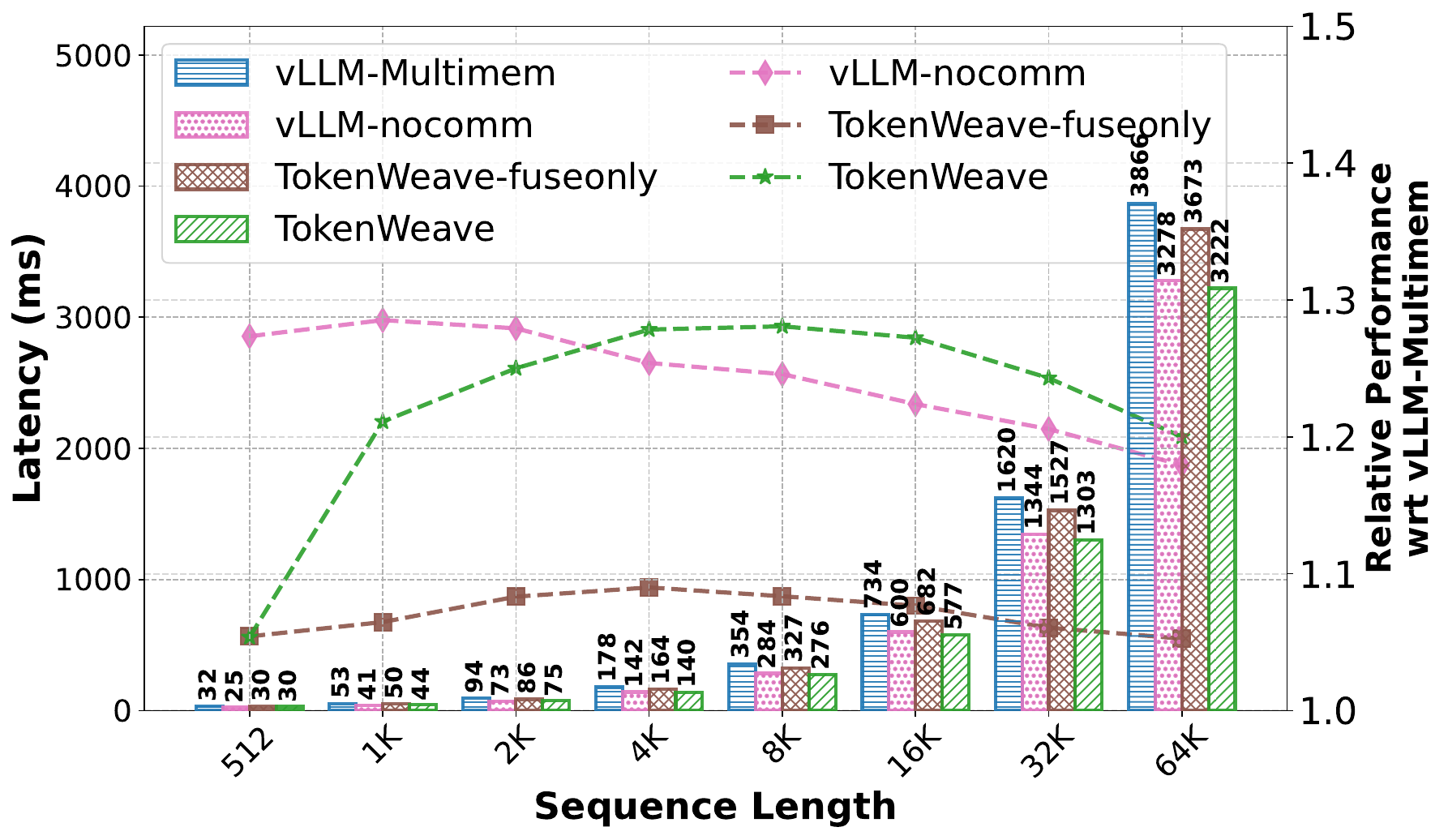}
    \vspace{-18pt}
    \caption{Inference latency of Llama-3.3-70B on an $8\times$H100 DGX system for various sequence lengths. \textit{vLLM-Multimem} corresponds to vLLM with an optimized AllReduce implementation using Multimem~\cite{MULTIMEM_PTX} and NVSHARP~\cite{NVSHARP} support. \textit{vLLM-nocomm} is a counterfactual baseline corresponding to only the computation time without any communication.
    The dotted lines show performance normalized to the \textit{vLLM-Multimem} baseline. \sysname achieves up to $1.28\times$ speedup. Even at shorter sequence lengths, \sysname provides significant gains, e.g., $1.2\times$ at a sequence length of $1K$ tokens.
    At sequence lengths $\ge 4K$, \sysname outperforms \textit{vLLM-nocomm} by not only recovering all communication overhead but also providing additional gains due to its AllReduce--RMSNorm fused kernel.}
\label{fig:cover-llama3-gain-comparison}
\end{figure}

In this paper, we present \sysname, a system that reduces communication overhead during distributed LLM inference. To the best of our knowledge, it is the first to mitigate tensor-parallel communication overheads in low-latency inference settings, delivering $1.2\times$ relative latency improvement even in iterations with as little as 1024 tokens~(\autoref{fig:cover-llama3-gain-comparison}), thus accelerating LLM serving.

\sysname comprises three key ideas, as follows. Our first insight is that RMSNorm, an operation that has previously been ignored as insignificant~\cite{nanoflow}, is crucial for optimizing tensor-parallel communication overheads. This is not only because RMSNorm has a non-trivial overhead of $4\text{--}9$\% (\autoref{fig:allreduce-rmsnorm-cost}), but also because the RMSNorm operation is intricately connected with the communication operation. vLLM today implements RMSNorm after the AllReduce communication. Alternatively, one could break AllReduce into equivalent ReduceScatter and AllGather operations and perform RMSNorm after the ReduceScatter operation on only the tensor shard local to each GPU. While this keeps the mathematical operation unchanged, it reduces the per-GPU RMSNorm computation by a factor equal to the tensor-parallel degree. However, this alternative is not used today because splitting AllReduce into ReduceScatter and AllGather results in significant overheads, as shown in \autoref{fig:allreduce-split-perf}, which outweigh any cost reduction in RMSNorm (\autoref{tab:fused-ar-rmsnorm}). In this paper, we show that an optimized AllReduce--RMSNorm fused kernel that carefully fuses ReduceScatter, RMSNorm and AllGather by minimizing HBM traffic can be significantly more efficient than either of the above alternatives. Further, this fused kernel also serves as a foundation to unlock compute-communication overlap for small token batches as discussed next.

\begin{figure*}[t]
    \centering
    \includegraphics[width=0.99\linewidth]{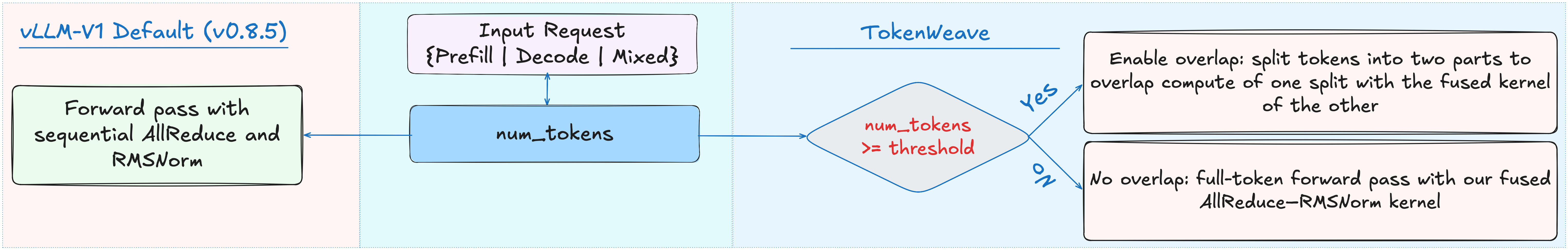}
    \vspace{-6pt}
    \caption{Selective enabling of splitting/overlap in \sysname. At each iteration, the \texttt{num\_tokens} in the batch is checked against a token threshold. Full \sysname with splitting and overlap is enabled only for higher values of \texttt{num\_tokens}. For smaller \texttt{num\_tokens}, where splitting can result in higher overheads, we only enable the fused AllReduce--RMSNorm kernel. The method applies uniformly to prefill-only, mixed, and decode-only batches.}
\vspace{-6pt}
\label{fig:sysname.drawio}
\end{figure*}

Second, \sysname{} overlaps communication with computation using at most two-way GPU-wave-aware token splits, compatible with both chunked-prefills and prefill/decode batches (\autoref{fig:sysname-schematic}). This allows communication (i.e., AllReduce) and RMSNorm (due to the fused kernel) of one split to proceed in parallel with the computation of the other. The splitting is carefully chosen to be wave-aware, ensuring that the total number of waves executed in kernels of the two splits is not more than the number of waves if there were a single non-split kernel, thus largely eliminating the compute overheads due to work splitting. For really small token batches, where splitting overheads remain, \sysname{} eschews splitting and simply uses the AllReduce--RMSNorm fused kernel (see \autoref{fig:sysname.drawio}).

Third, by fusing communication and normalization and leveraging modern hardware features such as Multimem~\cite{pytorch-multimem-allreduce}, \sysname{} achieves efficient overlap while reserving only $2\text{--}8$ SMs on H100 for both communication and normalization compared to $16\text{--}20+$ SMs used for communication alone in prior work~\cite{nanoflow, tbo}. \sysname{}'s fused kernel ensures that the memory-bandwidth-bound communication and normalization operations do not interfere with each other while Multimem minimizes communication SMs, ensuring that the bulk of the SMs are dedicated to computation. 

Our AllReduce--RMSNorm fused kernel shows consistent $1.34\text{--}1.39\times$ gains across token sizes from $64$ to $32K$ (\autoref{tab:fused-ar-rmsnorm}), almost matching the performance of standalone AllReduce. Thus, even in disaggregated settings~\cite{distserv, splitwise} where prefill and decode are isolated from each other, small decode-only batches directly benefit from kernel fusion, while larger prefill-only batches achieve even greater improvements when combined with our compute-communication overlap strategy, outperforming prior overlap approaches.

We integrate \sysname into the latest vLLM-V1 serving framework and evaluate its performance across various models. \autoref{fig:cover-llama3-gain-comparison} shows that \sysname achieves up to $1.28\times$ better overall latency compared to an optimized baseline, with $1.2\times$ latency gains even for as little as $1K$ tokens in the batch. In comparison, a state-of-the-art communication overlap solution, TileLink, results in a latency increase even at $2K$ tokens and delivers maximum gains of $1.2\times$ at $8K+$ tokens (\autoref{fig:tilelink-vs-tokenblend}). Further, for realistic end-to-end workloads, we find that \sysname also delivers up to $1.19\times$ and $1.15\times$ higher throughput for ShareGPT~\cite{sharegpt_vicuna_unfiltered} and arXiv~\cite{arxiv-summarization, arxiv_summarization_hf} traces, respectively (\autoref{fig:sysname-end-to-end-perf-hybrid}). Finally, in several settings, \sysname{} is even able to outperform an equivalent model with all communication removed since it also optimizes and overlaps the memory-bandwidth-bound RMSNorm operation.

In summary, we make the following contributions:
\begin{enumerate}
    \item We identify RMSNorm as a crucial operation and optimize it via reordering and a novel, fused AllReduce--RMSNorm kernel that minimizes HBM traffic.
    \item We design a wave-aware, two-way token split compatible with both chunked-prefills and prefill/decode batches to overlap compute and communication even for small token batches.
    \item We reduce the SMs required for communication and normalization to 2--8 by leveraging NVSHARP/Multimem, enabling overlap without starving compute.
    \item We present the first system that overlaps compute-communication for LLM serving of tensor-parallel sharded models, delivering up to $1.19\times$ higher throughput in co-located settings. In disaggregated settings, our fused kernel helps optimize small decode-only batches, whereas large prefill-only batches can benefit from full overlap.
\end{enumerate}
\section{Background}
\label{sec:background}
\begin{figure*}[tb]
    \centering
    \includegraphics[width=0.82\textwidth]{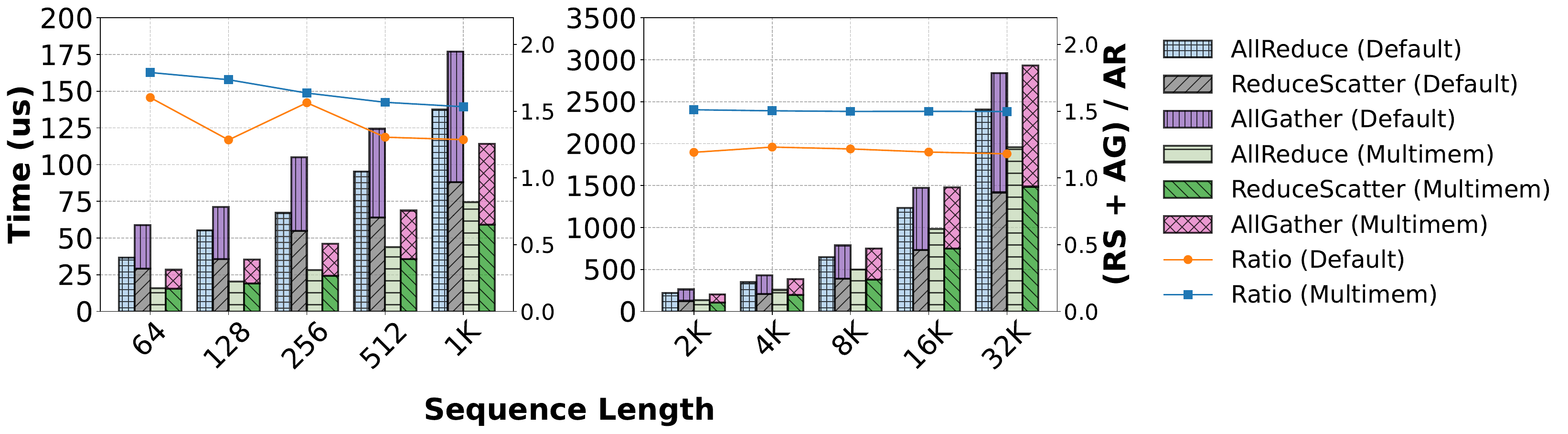}
    \vspace{-10pt}
    \caption{Splitting AllReduce (AR) into ReduceScatter (RS) and AllGather (AG) can result in non-trivial overheads. Shown are the absolute times and the relative performance (line plots) of these operations on an $8\times$H100 DGX system for varying sequence lengths. All runs are with a hidden size of $8192$ using $bf16$.}
\vspace{-6pt}
\label{fig:allreduce-split-perf}
\end{figure*}
We provide a brief overview of the operations performed during transformer model inference. A standard decoder-only Transformer~\cite{attentionpaper} processes an input sequence through multiple transformer blocks, each composed of a self-attention module, a feed-forward network (FFN), communication primitives (only in distributed settings), and normalization layers. The FFN layer typically consists of two linear transformations with a nonlinearity such as GELU~\cite{gelu} in between. The attention layer is composed of multiple attention heads. It first performs a \textit{QKV-}projection (i.e., pre-projection) operation to compute a \textit{query}, \textit{key}, and \textit{value} for each of the heads, which is followed by the self-attention operation in every head, and finally an \textit{O-}projection (i.e., post-projection) step that combines the output of all heads.

\subsection{Distributed Inference}
\label{sec:background:distributed}
The large size of many modern LLMs requires inference to be run over multiple GPUs in order to fit the model parameters. Even if the model fits on a single GPU, distributed inference may still be required to meet the strict latency service-level objectives (SLOs) of interactive workloads. Moreover, distributed inference is more efficient in many cases, as it allows for higher batch sizes by freeing up memory for the KV cache.

For intra-node distributed inference, the most common parallelism strategy is tensor parallelism (TP). In TP, the FFN weight matrices are partitioned across GPUs --- the first MLP is partitioned column-wise and the second row-wise~\cite{megatron}. Each GPU then performs computations using the local partial weight matrices. The individual outputs are then combined using an AllReduce operation 
to obtain the final output. For attention layers, the partitioning is done along the head dimension. Each GPU performs the \textit{QKV-}projection and self-attention for the local heads. The output of the final \textit{O-}projection matmul is finally combined across the GPUs using an AllReduce operation again.
\begin{figure}[tb]
    \centering
    \includegraphics[width=0.9\linewidth]{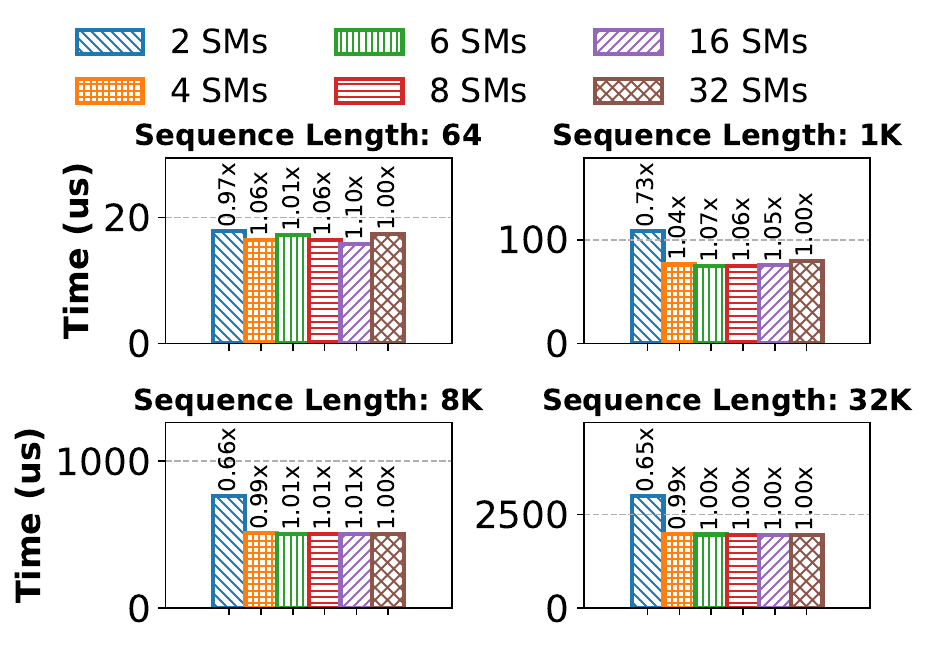}
    \vspace{-6pt}
    \caption{Multimem-based AllReduce implementations require very few SMs. Shown is the performance of the Multimem AllReduce kernel with different numbers of SMs for varying sequence lengths (hidden size $8192$, $bf16$). In most cases, $4\text{--}8$ SMs are enough to achieve near-optimal performance.}
\vspace{-6pt}
\label{fig:multimem-ctas}
\end{figure}
TP thus requires two AllReduce operations per transformer block, which lie on the critical path. This can add significant cost to inference latency and reduce GPU efficiency. For example, as shown in \autoref{fig:allreduce-rmsnorm-cost}, the communication can add an overhead of up to $23$\% even on an $8\times$H100 DGX system, which has high-speed interconnects.

\subsection{NVSHARP, Multicast, and SymmetricMemory}
\label{sec:background:nvsharp-multicast-symmetricmemory}
The fourth generation (or later) NVSwitch systems (NVLink4) available in Hopper and Blackwell (or later) systems incorporate dedicated SHARP (Scalable Hierarchical Aggregation and Reduction Protocol) engines, termed NVSHARP or NVLS. NVLS enables GPUs to issue Multimem Parallel Thread Execution (PTX)\footnote{Parallel Thread Execution (PTX) is a virtual machine instruction set architecture that has been part of CUDA from its beginning.} store/load\_reduce instructions to multicast addresses. These instructions leverage the switch fabric to (i) duplicate packets to each subscribed GPU and (ii) perform an in-network reduction before forwarding the aggregated result. Because arithmetic operations are executed directly within the switch ASIC, communication collectives significantly reduce both NVLink bandwidth usage and GPU SM resource consumption. Modern NVIDIA architectures such as Hopper and Blackwell include native support for NVSHARP/Multimem. We expect such support to become standard across upcoming NVIDIA and potentially AMD GPUs.

\texttt{PyTorch} v2.6.0 exposes NVLS through its \textit{SymmetricMemory} API~\cite{pytorchsymmetricmemory}. SymmetricMemory facilitates allocating peer buffers on GPUs using the \verb!symm_mem.empty()! call (similar to \verb!torch.empty()!). After buffer allocation, a collective call to \verb!symm_mem.rendezvous()! exchanges memory handles, mapping peer buffers into the virtual address space of each participating GPU. Post-rendezvous, each GPU can access remote or multicast pointers using standard memory operations within Triton or CUDA kernels, eliminating explicit NCCL calls and substantially simplifying the implementation of communication routines.

These hardware and software advancements considerably decrease the number of SMs required for executing communication primitives and alleviate memory bandwidth pressure. In our experiments, we observed that utilizing only $6\text{--}8$\% of SMs on H100 GPUs suffices to saturate communication bandwidth, leaving the majority of SMs free for compute tasks that overlap communication. This can be seen in \autoref{fig:multimem-ctas}, which shows the latency of the AllReduce kernel vs. the number of SMs.
\section{Related Work}
\label{sec:related-work}
A common strategy for reducing communication overhead is to overlap the communication with some other computation. However, because of data dependencies, the data to be communicated becomes available only after the computation steps (FFN/Attention) finish. To solve this, one approach is to break the computation into smaller subtasks. The communication of completed subtasks can then be overlapped with the computation of the remaining subtasks. Breaking into subtasks, however, can result in reduced compute efficiency. Modern GPUs are much more efficient at running large computation kernels due to the high degree of parallelism they offer, and breaking into smaller kernels can result in wave quantization effects where a bunch of GPU SMs may not have any work in the last computation wave. Many techniques~\cite{coconet, googleoverlap, async-tp, flux, tritondistributed, tilelink} have been proposed to fix this via fused kernel implementations, where the communication is orchestrated from within the compute kernels as the computation of individual tiles finishes. 

\citet{googleoverlap}, for example, break AllReduce into a ReduceScatter and an AllGather operation. The ReduceScatter is overlapped with the preceding computation, while the AllGather is overlapped with the subsequent computation.
To orchestrate the overlaps, they rely on XLA~\cite{xla} and TPU support to invoke async collective APIs from within a tiled GEMM kernel. The reliance on XLA, however, has made porting to PyTorch + CUDA difficult, resulting in limited mainstream adoption.
\begin{figure}[!tb]
    \centering
    \includegraphics[width=0.98\linewidth]{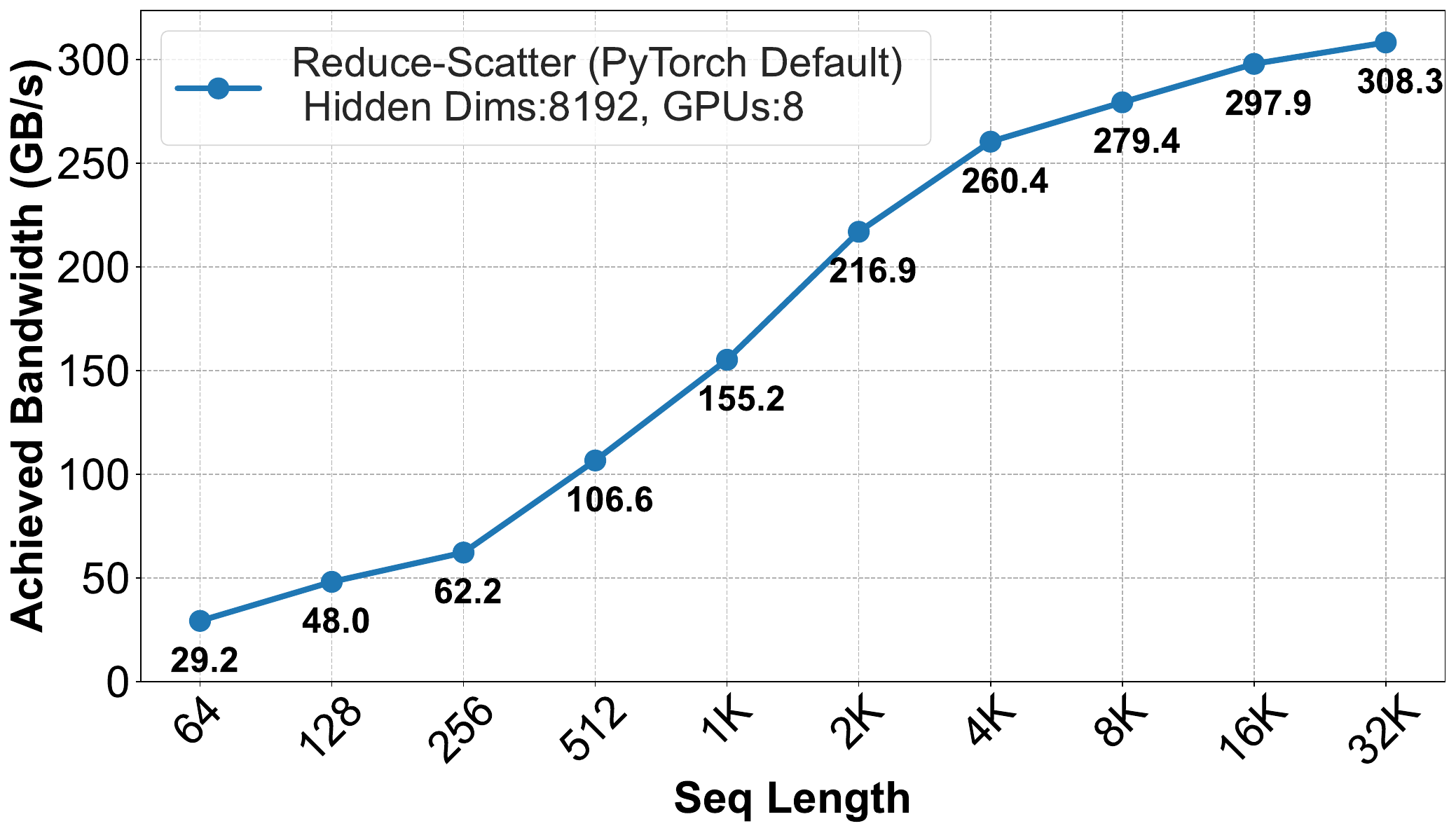}
    \vspace{-6pt}
    \caption{Large collective operations are more efficient. Shown is the bandwidth for ReduceScatter (RS) on an $8\times$H100 DGX system for varying sequence lengths (hidden size $8192$, $bf16$). Larger tensors result in much better bandwidth, demonstrating that splitting input into smaller parts results in overheads.}
\label{fig:reducescatter-vs-size}
\end{figure}

Flux~\cite{flux} provides a solution similar to that of~\citet{googleoverlap} for CUDA-based implementations. They also break AllReduce into ReduceScatter and AllGather, and take a fused kernel approach via a CTA-level streaming scheme. Flux decomposes computation and communication into fine-grained operations and fuses them into a single kernel that interleaves compute and communication within CTAs. Tiles that require remote operands fetch them via GPU-initiated communication (e.g., NVLink/NVSwitch P2P or NVSHMEM~\cite{NVSHMEM}), with transfers issued ahead of use so that independent tiles can proceed while data is in flight. This exposes opportunities to overlap communication with tensor-core MMA. However, the design of Flux and TileLink is fundamentally constrained by the gap between HBM and interconnect (i.e., NVLink) bandwidth: when communication is not fully hidden, remote accesses become throughput-limiting and fall on the critical path. In addition, tightly coupling communication with tile execution restricts scheduling flexibility, making performance sensitive to the compute-communication balance and less robust to workload variation.

The fused kernel approaches~\cite{googleoverlap, flux, tilelink, hassani2024distributedgemm, async-tp} suffer from three problems. First, since the techniques rely on overlap within a GEMM kernel, communication overlap during non-GEMM operations like attention is not feasible. For example, the AllReduce after the FFN layer is performed by overlapping the ReduceScatter portion with the second MLP, and overlapping the AllGather portion with the \textit{QKV-}projection of the next attention layer. 
This limits the opportunity for overlap. For small models or small batch sizes where the \textit{QKV-}projection takes a very small time, there is not enough time to overlap the AllGather. Similarly, the \textit{O-}projection does not take long enough to overlap the ReduceScatter. Second, the splitting into ReduceScatter and AllGather operations is less efficient compared to the equivalent AllReduce operation. \autoref{fig:allreduce-split-perf} plots the performance of a single AllReduce kernel compared to the cost of the equivalent ReduceScatter+AllGather for different communication sizes. As shown, this can add significant overheads, upwards of $50$\%. Third, the smaller tile-sized granularity of communication is less efficient than doing single large transfers. \autoref{fig:reducescatter-vs-size} plots the achieved ReduceScatter bandwidth vs the tensor size. As shown, smaller communication sizes achieve much lower bandwidths. Due to these factors, \citet{flux} and \citet{tilelink} are able to effectively hide communication only when the GEMM is large enough, requiring batches with large \tokensinbatch.
\begin{figure*}[t]
    \centering
    \begin{subfigure}{\textwidth}
        \centering
        \includegraphics[width=0.99\textwidth]{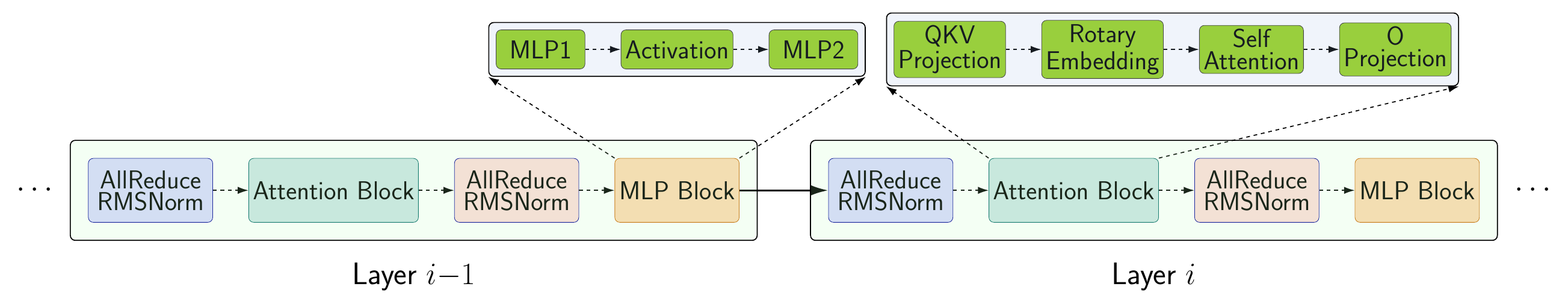}
        \caption{Vanilla Tensor Parallelism}
    \label{fig:tp-a}
    \end{subfigure}
    \begin{subfigure}{\textwidth}
        \centering
        \includegraphics[width=0.99\textwidth]{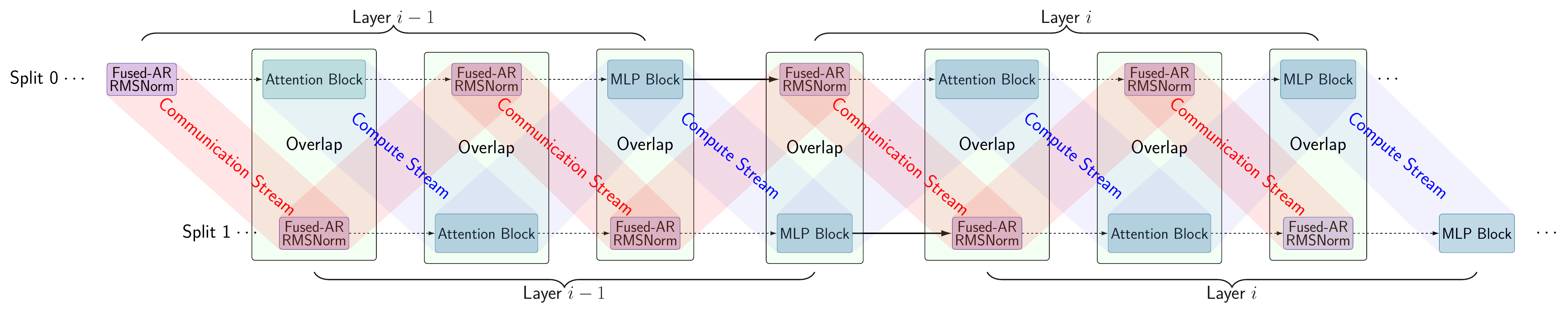}
        \caption{\sysname}
    \label{fig:tp-b}
    \end{subfigure}
    \caption{
    {\bf Overview of \sysname:} (a) Vanilla Tensor Parallelism: all compute and communication operations are performed sequentially. (b) \sysname: the input batch is partitioned into two splits. AllReduce is fused with RMSNorm, and computation of one split overlaps with communication of the other split. Separate compute and communication streams \textit{weave} to orchestrate the overlap.}
\label{fig:sysname-schematic}
\end{figure*}
\begin{figure}[t]
    \centering
    \includegraphics[width=1\linewidth]{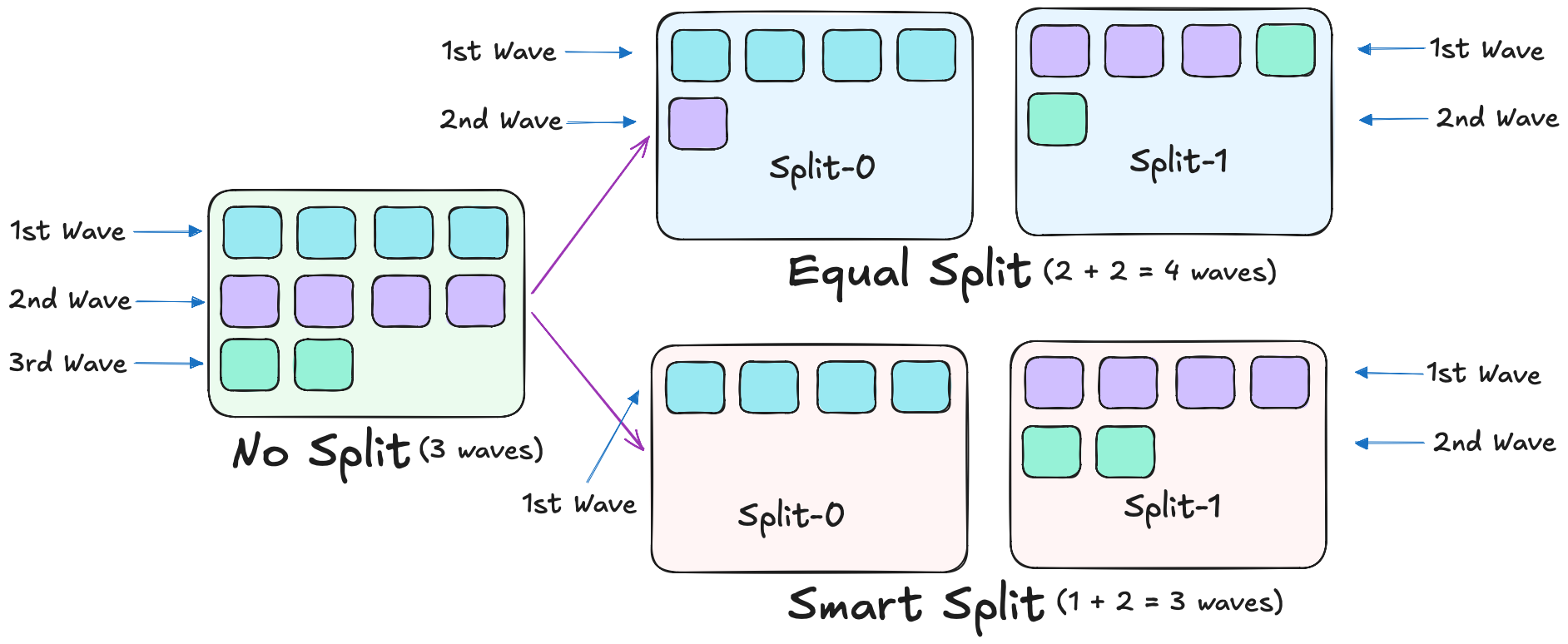}
    \vspace{-18pt}
    \caption{{\bf Benefits of \smartsplitting:} Execution of a kernel with ten CTAs on a hypothetical GPU with four SMs. Assuming each CTA exclusively occupies one SM, the kernel executes in multiple waves as SM resources become available.}
    \label{fig:smart-splitting.drawio}
\end{figure}

NanoFlow~\cite{nanoflow} attacks this problem from a scheduling angle: it slices an incoming batch into nano-batches --- at the granularity of whole kernels (FFN, attention, collective) rather than CTA tiles --- and assigns each nano-batch to a dedicated CUDA stream bound to a fixed subset of SMs. By co-scheduling nano-batches whose resource profiles complement one another (e.g., compute-heavy FFNs alongside communication-bound ReduceScatter), NanoFlow overlaps GPU compute, HBM traffic, and NVLink transfers. The approach, however, relies on high batch sizes for breaking the input batch into sufficiently sized nano-batches, as smaller kernels of nano-batches can result in significant overheads.

Splitting of batches has been proposed in multiple contexts in the past from training to inference~\cite{pipeline, pipemoe, lancet, nanoflow}. Most recently, DeepSeek uses the idea of batch splitting in their inference system~\cite{tbo} to achieve compute-communication overlap. However, their inference system uses expert parallelism, which needs expensive all-to-all communication instead of the cheaper AllReduce needed for tensor parallelism. They overlap this all-to-all communication of one batch with the computation of the second batch. Note that the communication cost in DeepSeek is $\sim50$\%~\cite{sglang-tbo}, compared to around $20$\% in TP (\autoref{fig:allreduce-rmsnorm-cost}). As a result, while splitting overheads can become a bottleneck when used with TP, they are not very critical for a DeepSeek type of setup, as the high communication overheads provide enough slack. Further, they do not overlap layer normalization along with communication.
\section{\sysname: Design and Implementation}
\label{sec:two-way-split}
We describe the techniques of \sysname --- smart \Tokensplitting, RMSNorm reordering, and a Multimem-based fused AllReduce--RMSNorm implementation. \autoref{fig:sysname-schematic} presents a high-level schematic comparing our method \textbf{(b)} to the standard tensor-parallel implementation \textbf{(a)}.

\sysname partitions the incoming batch into two subsets, each with nearly equal compute and communication requirements. \textbf{Why two subsets?} At least two subsets are needed to create a pipeline and avoid dependencies. Splitting into more than two segments is inefficient because it increases decomposition overhead without providing additional compute-communication overlap opportunities.

For illustration of the splitting approach, consider an incoming batch of size 3 with sequences comprising $L_1$, $L_2$, and $L_3$ tokens, respectively. These $T$ tokens ($T = L_1$ + $L_2$ + $L_3$) are divided into two parts: a \textit{prefix-split} containing the initial tokens of length $T_a$, and a \textit{suffix-split} containing the remaining subsequence of length $T_b$ ($T= T_a$ + $T_b$). These two splits, of lengths $T_a$ and $T_b$, are processed separately in a pipelined manner. All transformer operations, except attention, are token-level, and pose no problem due to the split processing. However, the attention operation introduces dependencies, as attention computations of tokens in the suffix subsequence depend on tokens in the prefix subsequence. To handle these dependencies, \sysname employs a chunked attention implementation~\cite{sarathi}, and ensures that operations for \textit{prefix-split} precede those of the \textit{suffix-split}. When processing batches larger than size 1, partitions may comprise either complete or partial sequences. \sysname{} ensures all prefixes of partial sequences reside within the \textit{prefix-split}, thereby preserving the necessary computational dependencies.
\begin{figure}[t]
    \centering
    \includegraphics[width=0.9\linewidth]{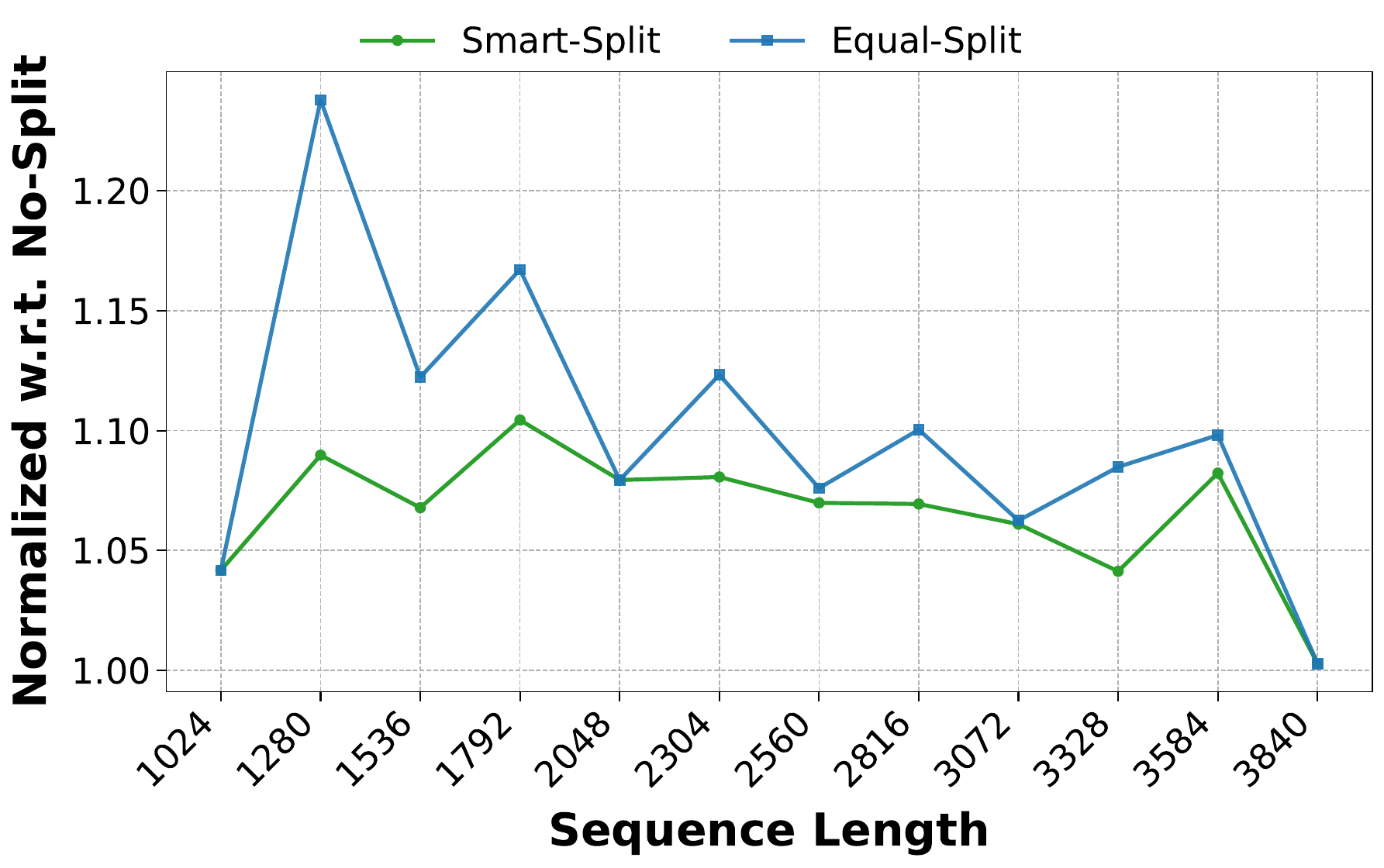}
    \vspace{-6pt}
    \caption{\Smartsplitting mitigates wave-quantization overheads. We compare \textit{smart-splits} with \textit{equal-splits} and show the normalized latency of the FFN layer w.r.t. \textit{no-split} case (Llama-3.3-70B on an $8\times$H100 DGX, $bf16$). \Smartsplitting reduces splitting overheads, especially for small sequence lengths.}
\label{fig:smart-splitting}
\end{figure}
\subsection{Coarse-Grained \Tokensplitting}
\label{sec:two-way-split:coarse-grained}
\subsubsection{Wave-Aware \Smartsplitting}
\label{sec:two-way-split:coarse-grained:smart-splitting}
Partitioning a large GPU computation into smaller units can introduce overhead due to wave quantization effects. Consider a GEMM kernel that requires 300 CTAs (Cooperative Thread Arrays). On an NVIDIA H100 GPU with 132 SMs (Streaming Multiprocessors), assuming each CTA occupies exactly one SM, this computation will span 2 full waves and 1 partial wave utilizing 36 SMs. The kernel's computation time will thus be $3\times$ the execution time of one wave. For clarity, \autoref{fig:smart-splitting.drawio} illustrates the wave-based execution of 10 CTAs on a hypothetical GPU with four SMs.

If this computation is evenly split into two batches of 150 CTAs each, each smaller batch now requires two waves: one full wave of 132 SMs followed by a partial wave of 18 SMs. Consequently, the two smaller computations collectively require four waves, thereby increasing execution time compared to the original unsplit computation.

To prevent such overhead, \Smartsplitting{} employs a wave-aware splitting strategy, ensuring that the combined waves required by both splits do not exceed the wave count of the original unsplit computation. In the example above, \Smartsplitting{} strategically divides the batch into one split containing 132 CTAs (exactly one full wave) and the other split with 168 CTAs (one full wave and one partial wave). This method effectively maintains the total computational waves, minimizing any wave quantization overheads due to splitting. \autoref{fig:smart-splitting} compares the latency of the FFN layer with and without \smartsplitting. As shown, \smartsplitting can reduce splitting overheads, especially for batches with fewer tokens. The \smartsplitting algorithm is shown in~\autoref{alg:smart_splitting_heuristics} in Appendix \textbf{A}.

\subsubsection{Overlapped Execution}
\label{sec:two-way-split:coarse-grained:overlapped-execution}
\begin{figure}[t]
    \centering
    \includegraphics[width=1\linewidth]{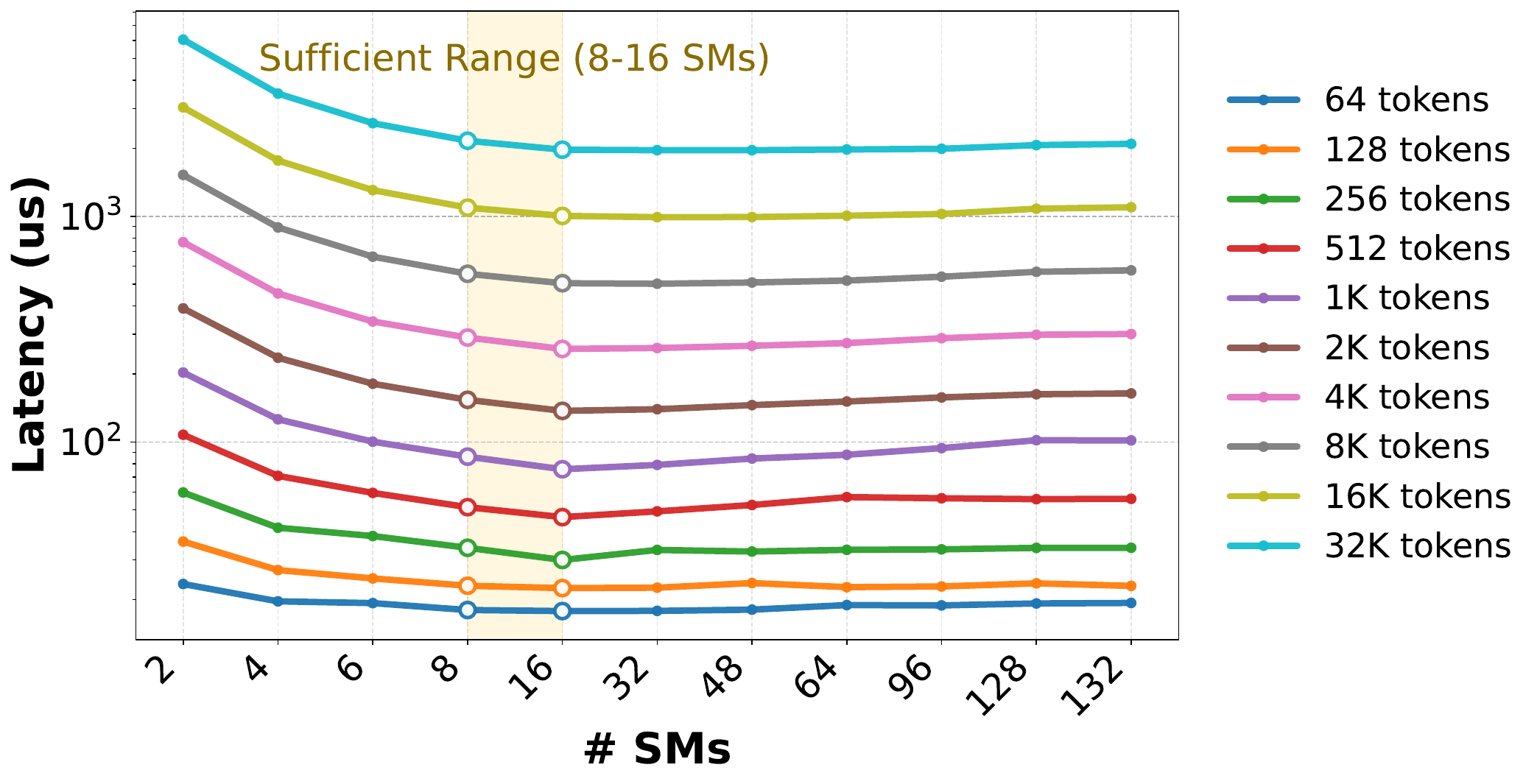}
    \vspace{-18pt}
    \caption{Our fused AllReduce--RMSNorm kernel performs optimally with very few SMs. We show the latency of the kernel under varying numbers of SMs for different sequence lengths (hidden size $8192$ on an $8\times$H100 DGX, $bf16$). Using $8$ SMs is close to optimal in most cases.}
\label{fig:allreduce-layernorm-fused-by-SM}
\end{figure}
The operations of these two batches can now be overlapped, as shown in \autoref{fig:tp-b}. As illustrated, when AllReduce (+RMSNorm because of the fused kernel) of the first batch is being processed, we compute the attention of the second batch. The FFN of the first batch is then overlapped with AllReduce (+RMSNorm) of the second batch and so on. We implement this overlapped execution via CUDA streams. A communication stream handles all communication operations, while a compute stream runs the compute operations. Lightweight synchronization using \texttt{torch.cuda.stream\_wait(stream, wait\_stream)} is performed between the compute and communication streams to handle data dependencies.

\subsection{RMSNorm Reordering}
\label{sec:two-way-split:rmsnorm-reordering}
\begin{table*}[t]
\centering
\small
\setlength{\tabcolsep}{6pt}
\renewcommand{\arraystretch}{1.30}
\resizebox{\linewidth}{!}{%
\begin{tabular}{lcccccccccc}
\toprule
\rowcolor{headercolor}
 \# Tokens & 64 & 128 & 256 & 512 & 1K & 2K & 4K & 8K & 16K & 32K \\
\midrule
\rowcolor{lowcolor}
AR 
& 16.32 & 20.64 & 28.35 & 43.84 
& 74.85 & 136.00 & 257.47 & 500.54 
& 986.24 & 1955.71 \\

\rowcolor{lowcolor}
RMSNorm 
& 8.32 & 9.57 & 12.06 & 18.91 
& 29.82 & 52.16 & 96.29 & 185.09 
& 361.54 & 716.13 \\
\hdashline
\rowcolor{midcolor}
AR+RMSNorm 
& 24.64 & 30.21 & 40.41 & 62.75 
& 104.67 & 188.16 & 353.76 & 685.63 
& 1347.78 & 2671.84 \\
\hdashline
\rowcolor{highcolor}
Simple Fusion
& 24.74 (\textbf{1.00}) & 30.11 (\textbf{1.00}) & 40.58 (\textbf{1.00}) & 63.94 (\textbf{0.98})
& 108.70 (\textbf{0.96}) & 201.66 (\textbf{0.93}) & 386.14 (\textbf{0.92}) & 715.33 (\textbf{0.96})
& 1345.95 (\textbf{1.00}) & 2495.10 (\textbf{1.07}) \\
\hdashline
\rowcolor{highcolor}
\textbf{Fused (Ours)} 
& 17.70 (\textbf{1.39}) & 22.53 (\textbf{1.34}) & 30.02 (\textbf{1.35}) & 46.46 (\textbf{1.35})
& 75.71 (\textbf{1.38}) & 137.34 (\textbf{1.37}) & 258.34 (\textbf{1.37}) & 502.24 (\textbf{1.37})
& 990.59 (\textbf{1.36}) & 1960.90 (\textbf{1.36}) \\
\bottomrule
\end{tabular}
}
\vspace{-8pt}
\caption{{\bf Fused AllReduce--RMSNorm kernel performance.} We show the latency of AllReduce and RMSNorm operations separately and the latency of both operations when performed sequentially (AR+RMSNorm), with simple fusion (RS+RMSNorm+AG) and with our fused kernel (hidden size $8192$, $us$, $bf16$, $8\times$H100 DGX). The numbers in parentheses show the relative performance compared to sequential (AR+RMSNorm) computation.
Simple fusion results in worse performance for token sizes from $512$ to $8K$ due to the overhead of splitting AR into RS and AG, as well as the different parallelization requirements of the communication and RMSNorm kernels. Our fused kernel results in significant gains of $1.34\text{--}1.39\times$ across the token range and almost matches the AR-only performance.}
\label{tab:fused-ar-rmsnorm}
\vspace{-12pt}
\end{table*}
Distributed inference via tensor parallelism involves an AllReduce operation after attention and FFN computations (see \S\ref{sec:background:distributed} for background on distributed inference). This is followed by the residual addition and RMSNorm operations, which are computed independently by each GPU and are generally fused in modern implementations. However, since all GPUs have identical token embeddings after AllReduce, this results in redundant computation.

To address this inefficiency, \sysname{} strategically reorders RMSNorm operations within the AllReduce process. Specifically, the AllReduce operation can be decomposed into ReduceScatter and AllGather operations. At the completion of the ReduceScatter step, each GPU possesses the complete and final state of $\frac{1}{N}$ of the tensor, where $N$ denotes the total number of GPUs in the distributed replica. Consequently, each GPU can independently perform RMSNorm on its dedicated $\frac{1}{N}$ portion without redundancy. Since RMSNorm is a token-level operation, we only need to ensure that ReduceScatter splits the tensor only at token boundaries. This ensures that each GPU has the full token embeddings. The subsequent AllGather operation then distributes these post-RMSNorm values to all GPUs.

Through this reordering, \sysname{} reduces the RMSNorm computation by a factor of $N$ compared to traditional implementations, thereby eliminating unnecessary redundancy. However, as shown in \autoref{tab:fused-ar-rmsnorm}, such a simple reordering can actually result in performance loss, as the cost of dividing AllReduce into separate ReduceScatter and AllGather cancels the RMSNorm computation gains. 

We analyzed the cause of the significant overhead of splitting AllReduce and found that the key difference between the integrated AllReduce implementation and the split implementation is the extra HBM reads and writes in the split implementation. While simple fusion fuses all three kernels into one, it is unable to reduce the HBM traffic overhead. Further, in some cases, simple fusion performs worse than the non-fused version because the optimal parallelization strategy for ReduceScatter or AllGather can be different from the optimal parallelization for RMSNorm, which only the non-fused version is able to exploit. We address these inefficiencies through a custom fused kernel, discussed next.

\begin{figure*}[t]
    \centering
    \includegraphics[width=0.96\textwidth]{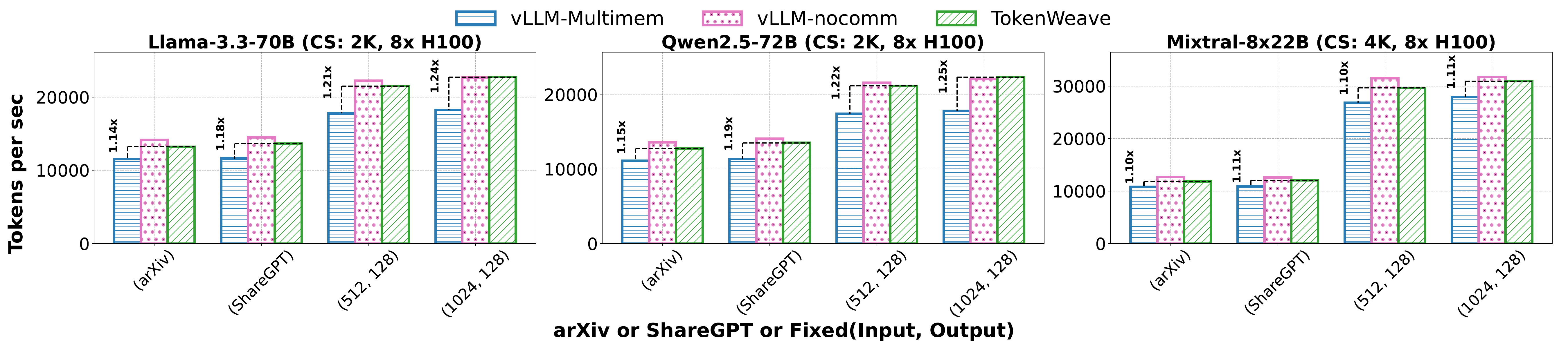}
    \vspace{-6pt}
    \caption{{\bf \sysname throughput gains for end-to-end workload traces.} We show the measured throughput for ShareGPT, arXiv, as well as fixed (input, output)-length traces for different models on $8\times$H100 DGX.}
\label{fig:sysname-end-to-end-perf-hybrid}
\end{figure*}

\begin{figure*}[t]
    \centering
    \includegraphics[width=\linewidth]{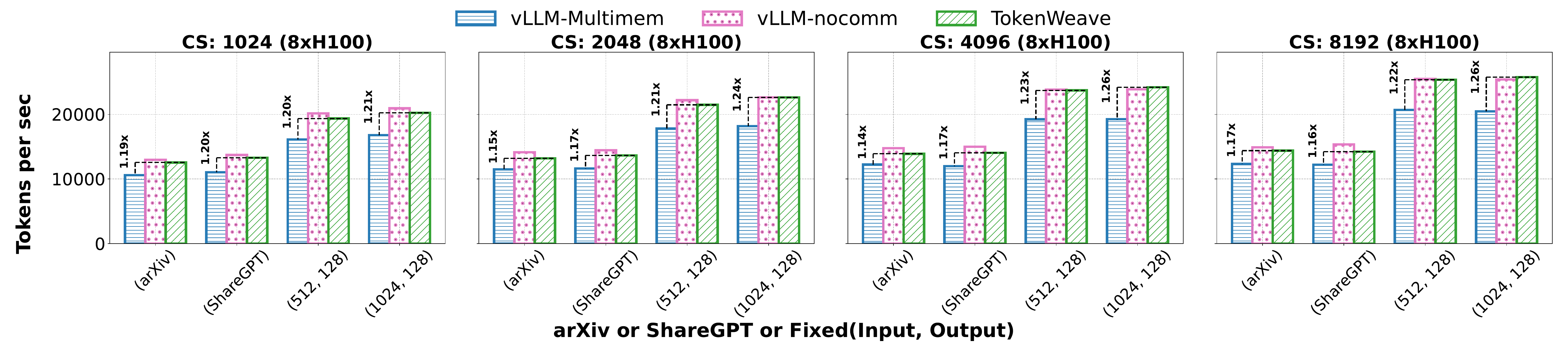}
    \vspace{-16pt}
    \caption{{\bf \sysname throughput gains for end-to-end traces under chunk size variation.} We show the measured throughput for ShareGPT, arXiv, as well as fixed (input, output)-length traces for Llama-3.3-70B on $8\times$H100 DGX. Chunk size varied from $1024\text{--}8192$.}
\vspace{-6pt}
\label{fig:sysname-llama-multiple-cs}
\end{figure*}

\subsection{Fused AllReduce--RMSNorm Implementation}
\label{subsec:fused-allreduce-rmsnorm-impl}
\sysname{} leverages Multimem capabilities (available on Hopper, Blackwell and future NVIDIA architectures) for an efficient, fused implementation of ReduceScatter, RMSNorm, and AllGather operations. At a high level, during ReduceScatter, each GPU uses NVSHARP to instruct the switch to perform the reduction of its $\frac{1}{N}$-th portion of the tensor. We then perform RMSNorm computation immediately on this reduced portion available in the SM registers without using the HBM, and the results are subsequently directly written to the Multimem addresses for the AllGather distribution by the switch to all GPUs.

We provide the source code of our fused kernel in \autoref{lst:fused-ar-rmsnorm-pseudo-code} (Appendix \textbf{A}). A standard RMSNorm typically necessitates two HBM reads over the entire tensor --- one to compute the variance of the token embeddings and another to scale the values by the computed variance --- as well as one final HBM write, again over the whole tensor. In contrast, our fused implementation operates only on the local shard assigned to each GPU, rather than on the full tensor, and optimizes memory access by computing the variance (line 25: Sum of squares) directly on the result of the Multimem ReduceScatter (line 23), thereby eliminating the initial HBM read. Additionally, we save an extra HBM write by directly outputting the normalized values to Multimem for the AllGather operation (line 36). Further, note that the residual addition operation is also fused with RMSNorm (line 24). 

The reduction in memory accesses, combined with the elimination of redundant computations when RMSNorm is performed after AllReduce, makes our fused AllReduce--RMSNorm kernel very efficient. As shown in \autoref{tab:fused-ar-rmsnorm}, the fused kernel provides up to $1.39\times$ improvement over the current approach of separate AllReduce + RMSNorm computation across all sequence lengths on an $8\times$H100 DGX.

Also, the reduced compute and memory bandwidth requirements allow this fused kernel to be executed with very few SMs (only $2\text{--}8$ in our experiments on $8\times$H100) without incurring much overhead --- as shown in \autoref{fig:allreduce-layernorm-fused-by-SM}, the kernel execution time does not improve much after $8$ SMs. This allows us to overlap the full AllReduce--RMSNorm operation of one split-batch with the compute of the other split, enabling overlap of not just the communication but also the memory-bound RMSNorm operation. Since the fused kernel overlaps with computation from another split batch, allocating fewer SMs (e.g., four instead of eight) may incur some overhead; however, this overhead remains hidden as long as its execution time does not exceed that of the overlapped compute kernel.

\section{Experimental Evaluation}
\label{sec:experimental-evaluation}
We evaluate \sysname across a range of popular models and workload settings, and under multiple tensor parallelism (TP) configurations. In our evaluation, we seek to answer the following questions:
\begin{enumerate}[leftmargin=*]
    \item How much throughput improvement does \sysname provide when running real-world end-to-end workload traces? (\S\ref{sec:perf-eval:throughput}).
    \item How much communication latency overhead does \sysname recover for different models under different TP sizes and batch configurations? (\S\ref{sec:perf-eval:latency}).
    \item How does \sysname compare to prior approaches such as TileLink and NanoFlow?
    (\S\ref{sec:perf-eval:single-layer} and \S\ref{sec:perf-eval:nanoflow}).
\end{enumerate}

\subsection{Experimental Setup}
\label{sec:experimental-evaluation:experimental-setup}
\begin{figure*}[!tb]
    \centering
    \includegraphics[width=0.95\textwidth]{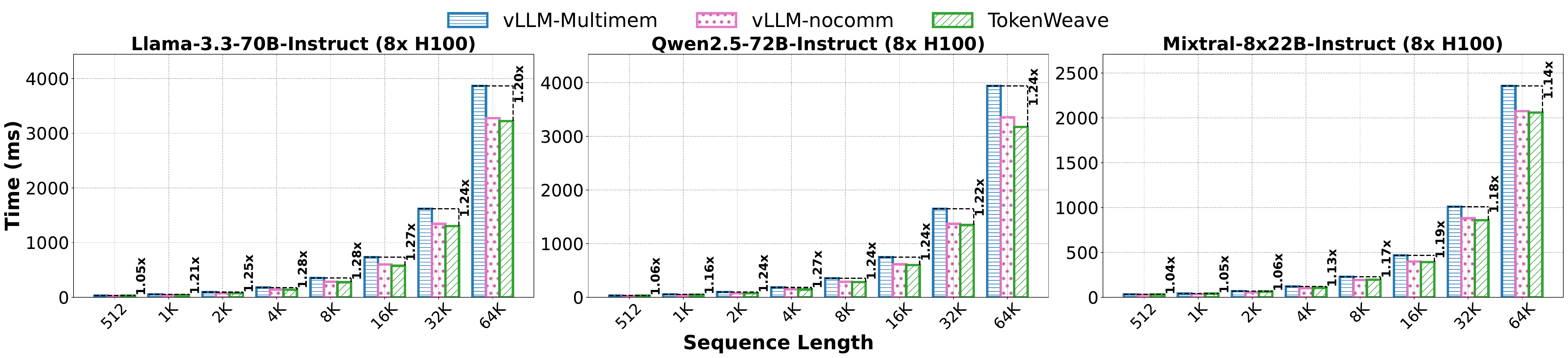}
    \vspace{-6pt}
    \caption{{\bf \sysname latency gains.} Execution times of prefill requests with varying sequence lengths for different models on $8\times$H100. \sysname is close to or better than the theoretical \textit{vLLM-nocomm} baseline with zero communication overhead, showing that \sysname not only recovers all communication overhead, but provides additional gains due to RMSNorm optimization.}
\label{fig:end-to-end-perf-prefill-latency-8h100}
\end{figure*}
\noindent\textbf{Implementation Details:} We implement \sysname on top of the \texttt{vLLM}~\cite{vllm} framework, a state-of-the-art serving system for LLM inference. Specifically, we build on the V1 engine~\cite{vllmv1} available in \texttt{vLLM}~0.8.5. Our implementation uses \texttt{PyTorch}~2.6.0 with \texttt{CUDA}~12.4. PyTorch provides support for \textit{symmetric memory}, which we use to implement communication collectives via \texttt{Triton}~3.2.0 and custom PyTorch CUDA extensions~\cite{pytorchsymmetricmemory}. In addition, we use the FlashAttention-3 backend for attention computation.

\noindent\textbf{Models:} We evaluate \sysname on three popular models, Llama-3.3-70B, Qwen2.5-72B and Mixtral-8x22B, that are served on multiple GPUs using tensor parallelism today. The first two are dense models, while Mixtral is a mixture-of-experts (MoE) model. 

\noindent\textbf{Workloads:} We evaluate over real-world traces, \textbf{ShareGPT}~\cite{sharegpt_vicuna_unfiltered} and \textbf{arXiv}~\cite{arxiv-summarization, arxiv_summarization_hf}, which have variable input prompt and output lengths. We also evaluate over synthetic fixed input and output length workloads to quantify the dependence on request length.

\noindent\textbf{Environment:} All our experiments in this section are performed on an $8\times$H100 NVIDIA DGX system with NVSHARP support, 128 CPU cores, and 800 GB of host memory. The \textbf{TP-4} experiments in Appendix \textbf{B} use the same system but utilize only four of the eight GPUs. Additional experimental results on an $8\times$B200 DGX system are available in Appendix \textbf{C}. We also use fixed TDP settings to ensure stable performance measurements~\cite{nvidia2022setstablepowerstate}. The CLI tool \lstinline{nvidia-smi} can be used to lock the GPU clocks to the base TDP frequency for the entire GPU, for example, using \lstinline{nvidia-smi --lock-gpu-clocks=tdp,tdp}. The clocks remain fixed until they are reset with \lstinline{nvidia-smi --reset-gpu-clocks}. For microbenchmarks, we perform warm-up, flush the L2 cache, and use \verb!cudaEvent!s for timing measurements. For end-to-end (e2e) results, we perform warm-up and then use \verb!time.perf_counter()! for timing. We disable prefix caching to ensure consistent and fair comparisons. 
For e2e evaluations, we ignore the detokenization time (time to convert output tokens back to text performed on the CPU). 
\begin{figure}[tb]
    \centering
    \includegraphics[width=0.9\linewidth]{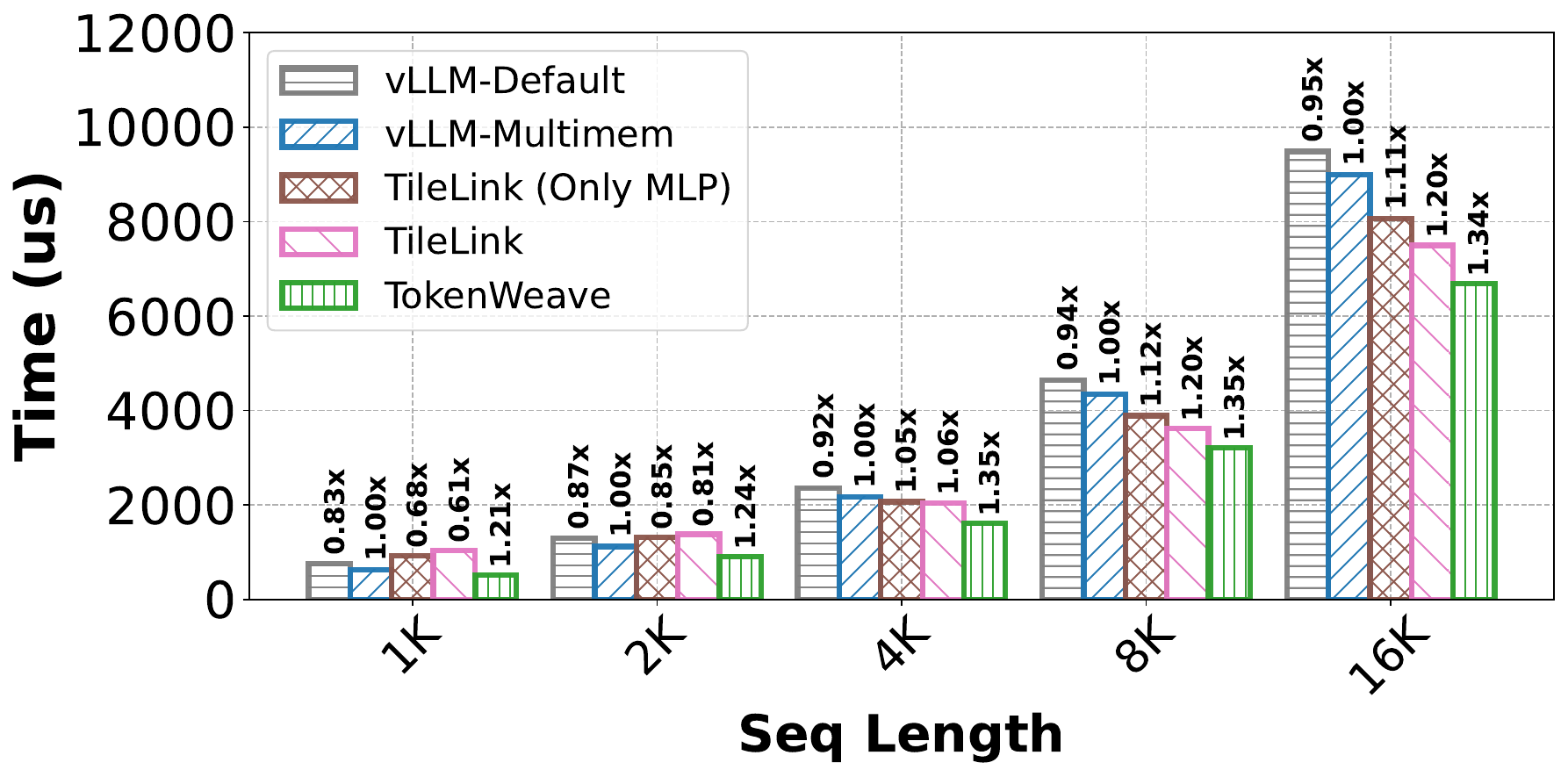}
    \vspace{-6pt}
    \caption{Single-layer latency for Llama-3.3-70B on an $8\times$H100 DGX system. Numbers at the top represent normalized performance compared to \textit{vLLM-Multimem}. While TileLink ends up with an overhead at small sequence lengths, \sysname consistently provides high gains over the entire sequence length range.}
\vspace{-6pt}
\label{fig:tilelink-vs-tokenblend}
\end{figure}

\noindent\textbf{Baselines:} We first compare against the \texttt{vLLM}~0.8.5 implementation that does not overlap communication operations. We evaluate both the default AllReduce implementation, called \textit{vLLM-Default}, and an optimized AllReduce implementation that leverages the NVSHARP and Multimem support on modern NVIDIA architectures, called \textit{vLLM-Multimem}. Further, we also report the performance of vLLM with all communication operations removed, called \textit{vLLM-nocomm}. While this version will not produce correct model output, it serves as a performance reference.

We also compare with TileLink~\cite{tilelink} which is the current state-of-the-art method for reducing communication overheads using compute-communication fusion and outperforms Flux~\cite{flux}, a technique that optimizes compute, memory bandwidth and communication. However, note that TileLink\footnote{\scriptsize\url{https://github.com/ByteDance-Seed/Triton-distributed/tree/b10fa2e}}
is not integrated with vLLM, which makes end-to-end comparison difficult. We thus use an optimized single-layer implementation of Llama-3.3-70B to compare \sysname against TileLink. NanoFlow again has its own serving stack and is not integrated with vLLM. Thus, to compare against the NanoFlow\footnote{\scriptsize\url{https://github.com/efeslab/Nanoflow/tree/e2e3c2a}}
baseline, we look at the communication overhead that is recovered in NanoFlow against their baseline framework vs. that recovered by \sysname against vLLM, which is our baseline framework. We also report absolute numbers for reference.

\subsection{Experimental Results}
\label{sec:perf-eval}
We now present our experimental findings. We first evaluate \sysname's end-to-end performance and then compare it against TileLink and NanoFlow.

\subsubsection{\sysname Throughput Gains}
\label{sec:perf-eval:throughput}

We first evaluate the throughput performance of \sysname for various workload traces. As is standard practice, we use chunked-prefills~\cite{sarathi} with hybrid prefill--decode batching to ensure consistent latency guarantees while maximizing throughput. Chunked-prefills ensures that time-between-tokens (TBT) latency is bounded, resulting in a smooth response to the user. Chunked-prefills is turned on by default in the vLLM-V1 framework. For the first experiment, we use a chunk size of $2K$ (\texttt{vLLM}~0.8.5 default) for the dense models and $4K$ for Mixtral. We apply \sysname to hybrid batches with $1K$ or more tokens ($4K$ for Mixtral) and use non-overlapped communication via our fused kernel for the smaller decode-only batches.

\autoref{fig:sysname-end-to-end-perf-hybrid} shows the performance evaluation with chunked-prefills on real-world workloads, such as ShareGPT and arXiv. These datasets contain requests with varying input and output lengths. To evaluate \sysname's dependence on request lengths, we also present performance results on fixed input and output length requests. As shown, in the $8\times$H100 setting, \sysname continues to show substantial throughput gains of approximately $1.19\times$ for dense models, effectively recovering most of the communication overhead. \autoref{fig:sysname-end-to-end-perf-hybrid-4gpus} in the Appendix shows the corresponding results for the 4-GPU setting, again demonstrating strong performance improvements.

The chunk size used in chunked-prefills enables a trade-off between TBT latency and throughput, with smaller chunk sizes enabling lower TBT values but also resulting in lower throughput. In the second experiment, we vary the chunk size and evaluate \sysname's throughput.
\autoref{fig:sysname-llama-multiple-cs} uses the same workload as before but with varying chunked-prefills sizes for the Llama-3.3-70B model. The figure shows consistent performance improvements of $1.14\times$ to $1.26\times$ with \sysname across a range of chunk sizes.

\subsubsection{\sysname Latency Benefits}
\label{sec:perf-eval:latency}
\begin{figure}[t]
    \centering
    \includegraphics[width=0.95\linewidth]{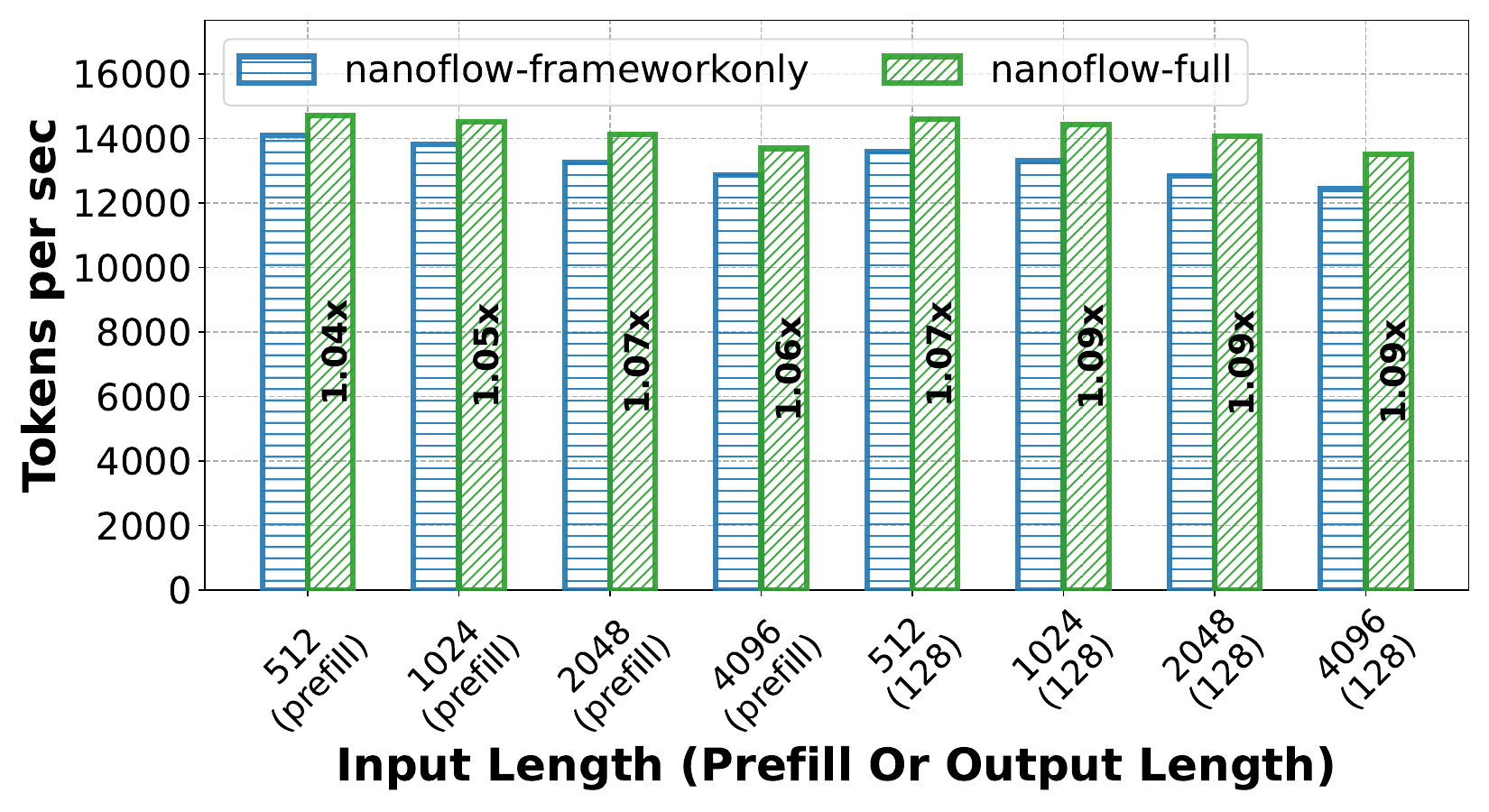}
    \vspace{-11pt}
    \caption{NanoFlow throughput for end-to-end workload traces under fixed (input, output)-length traces for Llama-3.3-70B on an $8\times$H100 DGX. \textit{nanoflow-full} corresponds to full NanoFlow, while \textit{nanoflow-frameworkonly} disables NanoFlow (both nanobatching and overlap) but uses their custom serving framework.}
\label{fig:nanoflow}
\end{figure}
\begin{figure*}[t]
    \centering
    \includegraphics[width=0.95\textwidth]{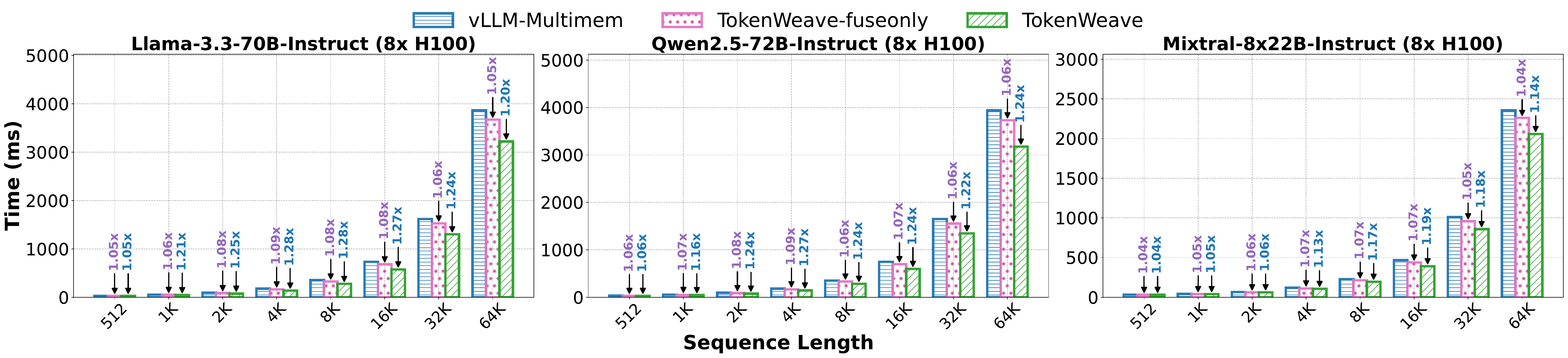}
    \vspace{-6pt}
    \caption{We compare \textit{\sysname-fuseonly} and full \sysname against the \textit{vLLM-Multimem} baseline. Execution times are shown for prefill requests with varying sequence lengths for different models on $8\times$H100. \textit{\sysname-fuseonly} provides gains due to the elimination of redundancy in RMSNorm computation and intermediate memory accesses, while \sysname provides additional gains from the compute-communication overlap.}
\label{fig:sysname-end-to-end-perf-prefill-fuse-only}
\end{figure*}
We now evaluate latency for a single iteration of the forward pass. Results with prefill-only batches with varying sequence lengths are shown in \autoref{fig:end-to-end-perf-prefill-latency-8h100} and \autoref{fig:sysname-end-to-end-perf-prefill-4GPUs} (Appendix). We evaluate five different settings, corresponding to three models running on $8\times$H100 and $4\times$H100, except for Mixtral-8x22B on $4\times$H100 due to insufficient memory for the KV cache. \sysname consistently achieves between $1.16\times$ and $1.28\times$ improvement from $1K$ onward for dense models on $8\times$H100, with notable gains even at shorter sequence lengths. For the MoE Mixtral model, the gains are generally smaller, as the communication overhead for Mixtral itself is also lower (\autoref{fig:allreduce-rmsnorm-cost}). Further, unlike the dense models, there is a net overhead for the \sysname full implementation in Mixtral at smaller sequence lengths of $1K$ and $2K$. This is because, in MoE, the tokens for MLP computation get split across the experts ($8$ for Mixtral-8x22B), making the FFN computations memory-bound for smaller requests. As a result, even \sysname's coarse splitting into two subtasks can result in non-trivial overheads that cannot be compensated by savings from computation overlap. However, we still see gains at smaller sequence lengths ($1K$ and $2K$) by selectively enabling only the fused kernel, with no overlapping, as depicted in \autoref{fig:sysname.drawio}. For the $4\times$H100 experiments (\autoref{fig:sysname-end-to-end-perf-prefill-4GPUs} in the Appendix), \sysname again gets good gains consistently over a range of sequence lengths starting from $1K$. The gains are lower than $8\times$H100, as the communication overhead itself is less with fewer GPUs. Finally, we also show the latency numbers under varying batch sizes in \autoref{fig:sysname-end-to-end-perf-prefill-varying-batch} (Appendix), again showing substantial improvements.

\subsubsection{Comparison with TileLink}
\label{sec:perf-eval:single-layer}
\begin{figure*}[t]
    \centering
    \includegraphics[width=0.9\textwidth]{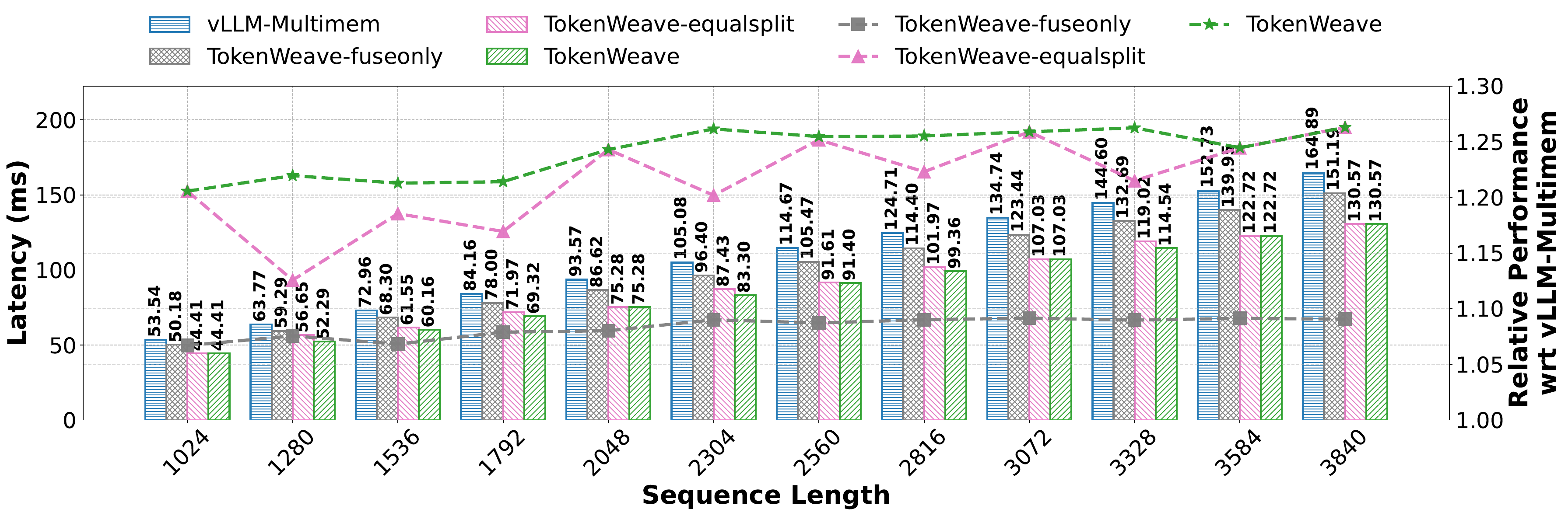}
    \vspace{-6pt}
    \caption{{\bf Ablation results.} \textit{\sysname-fuseonly} uses only the fused kernel. \textit{\sysname-equalsplit} enables token splitting and overlap but does not apply smart-splitting. Experiments are conducted on an $8\times$H100 DGX system.}
\label{fig:prefill_ablation_llama3_8xh100}
\end{figure*}
We now compare \sysname performance against TileLink, \textit{vLLM-Default} and \textit{vLLM-Multimem}. As noted earlier, we use a single layer of the model for comparison since TileLink is not integrated into a serving stack. We also compare against an additional baseline, TileLink-OnlyMLP, which performs the TileLink overlap only for the MLP layers. While this overlaps only one of the two communications per layer (i.e., AG of post-attention and RS of post-MLP), we find that it can be more efficient for smaller batches where the attention projection layers do not provide enough overlap opportunity and trying to overlap can increase rather than decrease overhead (recall that, unlike \sysname, TileLink can only overlap communication during GEMM computations). 

For this evaluation, we use batch size $1$ and vary the sequence length. \autoref{fig:tilelink-vs-tokenblend} shows the time taken for a single-layer computation. As shown, \sysname achieves a $1.20\times$ latency speedup even at a small request length of $1K$ tokens, while state-of-the-art TileLink actually ends up with a net overhead at small sequence lengths. TileLink results in performance improvements only for sequence lengths $\geq 4K$, but the improvements saturate at around $1.2\times$, while \sysname provides significantly higher improvements of up to $1.35\times$. Note that the end-to-end latency gains reported in \S\ref{sec:perf-eval:latency} are lower than the single-layer gains shown here due to additional non-layer overheads such as embedding, sampling, and other runtime components.

\subsubsection{Comparison with NanoFlow}
\label{sec:perf-eval:nanoflow}
NanoFlow~\cite{nanoflow} uses a custom serving stack and is not integrated with popular inference frameworks such as vLLM. This makes direct comparison of NanoFlow with \sysname difficult. Also, the NanoFlow implementation requires framework-level changes to co-ordinate the multiple nanobatches, which makes single-layer microbenchmarks impractical. Furthermore, NanoFlow's current implementation only supports NVIDIA A100 GPUs, whereas \sysname targets modern NVIDIA GPUs such as H100s and later. To facilitate a meaningful comparison, we adapted NanoFlow’s custom serving stack to run on H100 GPUs by implementing necessary modifications.

While the NanoFlow paper shows that their serving stack is more efficient than an earlier version of vLLM on A100 GPUs, we find our vLLM-V1 (0.8.5) baseline throughput numbers on H100 GPUs are better than those of NanoFlow's serving stack, perhaps because vLLM-V1 has significantly improved its performance recently. Thus, instead of performing an absolute throughput comparison, we resort to a relative performance comparison as described next.

We evaluate the adapted NanoFlow implementation under two conditions: with NanoFlow enabled and disabled within its own serving stack. \autoref{fig:nanoflow} shows the performance evaluation with chunked-prefills (chunk size $2K$) on workloads with various fixed input and output length requests. As shown, NanoFlow provides only modest performance gains in the range of $1.04\times$ to $1.09\times$, closely matching the $1.07\times$ communication improvement reported in the NanoFlow paper. In contrast, \sysname demonstrates significantly larger improvements of approximately $1.19\times$ across a broad range of scenarios (see \autoref{fig:sysname-end-to-end-perf-hybrid}).

\subsubsection{Ablation Study}
\label{sec:ablation-studies}
\autoref{fig:sysname-end-to-end-perf-prefill-fuse-only} shows ablation results for the fused AllReduce--RMSNorm kernel. \textit{\sysname-fuseonly} delivers speedups of $1.04\times\text{ to }1.09\times$ across all models due to the elimination of redundant computation and HBM accesses. Additionally, \sysname provides substantial gains over \textit{\sysname-fuseonly} due to the overlapped communication. The 4$\times$H100 configuration (\autoref{fig:sysname-end-to-end-perf-prefill-fuse-only-4h100} in the Appendix) shows similar results.

\autoref{fig:prefill_ablation_llama3_8xh100} shows an added ablation for \smartsplitting, comparing \sysname gains with and without \smartsplitting. Without \smartsplitting, the performance gains of \sysname see jitter depending on the exact sequence length. This is because the wave quantization effects from na\"ive splitting can be noticeable for some sequence lengths and negligible for others. \Smartsplitting removes this jitter almost completely, providing consistent gains throughout.
\section{Conclusion}
\label{sec:conclusion}
We find that the communication cost for large models served over multiple GPUs is as high as $20$\% today despite hardware support such as high-speed NVLink and NVSHARP. Furthermore, we identify that RMSNorm also results in an additional overhead of $4\text{--}9$\%. \sysname addresses these issues by splitting the model input into two approximately equal batches and overlapping the computation of one batch with the execution of a novel fused AllReduce--RMSNorm kernel on the other batch. Through extensive experimental evaluations using multiple models running on $8\times$H100, $4\times$H100, and $8\times$B200 DGX GPUs (see the Appendix for B200 and $4\times$H100 results), we show that \sysname achieves up to $1.28\times$ relative latency improvement and up to $1.19\times$ higher throughput compared to an optimized baseline in a variety of settings and workloads.

\bibliography{references}

\begin{thebibliography}{42}
\providecommand{\natexlab}[1]{#1}
\providecommand{\url}[1]{\texttt{#1}}
\expandafter\ifx\csname urlstyle\endcsname\relax
  \providecommand{\doi}[1]{doi: #1}\else
  \providecommand{\doi}{doi: \begingroup \urlstyle{rm}\Url}\fi

\bibitem[Agrawal et~al.(2024)Agrawal, Kedia, Panwar, Mohan, Kwatra, Gulavani, Tumanov, and Ramjee]{sarathi}
Agrawal, A., Kedia, N., Panwar, A., Mohan, J., Kwatra, N., Gulavani, B., Tumanov, A., and Ramjee, R.
\newblock Taming $\{$Throughput-Latency$\}$ tradeoff in $\{$LLM$\}$ inference with $\{$Sarathi-Serve$\}$.
\newblock In \emph{18th USENIX Symposium on Operating Systems Design and Implementation (OSDI 24)}, pp.\  117--134, 2024.

\bibitem[{arXiv}(2021)]{arxiv_summarization_hf}
{arXiv}.
\newblock arxiv summarization dataset.
\newblock \url{https://huggingface.co/datasets/ccdv/arxiv-summarization}, 2021.
\newblock Dataset.

\bibitem[Chang et~al.(2024)Chang, Bao, Hou, Jiang, Zheng, Zhong, Zhang, Song, Jiang, Lin, Jin, and Liu]{flux}
Chang, L.-W., Bao, W., Hou, Q., Jiang, C., Zheng, N., Zhong, Y., Zhang, X., Song, Z., Jiang, Z., Lin, H., Jin, X., and Liu, X.
\newblock Flux: Fast software-based communication overlap on gpus through kernel fusion, 2024.

\bibitem[Cohan et~al.(2018)Cohan, Dernoncourt, Kim, Bui, Kim, Chang, and Goharian]{arxiv-summarization}
Cohan, A., Dernoncourt, F., Kim, D.~S., Bui, T., Kim, S., Chang, W., and Goharian, N.
\newblock A discourse-aware attention model for abstractive summarization of long documents.
\newblock In \emph{Proceedings of the 2018 Conference of the North {A}merican Chapter of the Association for Computational Linguistics: Human Language Technologies, Volume 2 (Short Papers)}, pp.\  615--621, New Orleans, Louisiana, June 2018. Association for Computational Linguistics.
\newblock \doi{10.18653/v1/N18-2097}.
\newblock URL \url{https://aclanthology.org/N18-2097}.

\bibitem[{DeepSeek-AI}(2025)]{tbo}
{DeepSeek-AI}.
\newblock Profiling data in deepseek infra, 2025.
\newblock URL \url{https://github.com/deepseek-ai/profile-data?tab=readme-ov-file#inference}.

\bibitem[Grattafiori et~al.(2024)Grattafiori, Dubey, Jauhri, Pandey, Kadian, Al-Dahle, Letman, Mathur, Schelten, Vaughan, et~al.]{llama3}
Grattafiori, A., Dubey, A., Jauhri, A., Pandey, A., Kadian, A., Al-Dahle, A., Letman, A., Mathur, A., Schelten, A., Vaughan, A., et~al.
\newblock The llama 3 herd of models.
\newblock \emph{arXiv preprint arXiv:2407.21783}, 2024.

\bibitem[Hassani et~al.(2024)Hassani, Isaev, McDonald, Ren, Thakkar, Wu, and Shi]{hassani2024distributedgemm}
Hassani, A., Isaev, M., McDonald, N., Ren, J., Thakkar, V., Wu, H., and Shi, H.
\newblock Distributed gemm, 2024.
\newblock URL \url{https://blog.shi-labs.com/distributed-gemm-88be6a481e2b}.

\bibitem[Hendrycks \& Gimpel(2016)Hendrycks and Gimpel]{gelu}
Hendrycks, D. and Gimpel, K.
\newblock Gaussian error linear units (gelus), 2016.

\bibitem[Huang et~al.(2019)Huang, Cheng, Bapna, Firat, Chen, Chen, Lee, Ngiam, Le, Wu, et~al.]{pipeline}
Huang, Y., Cheng, Y., Bapna, A., Firat, O., Chen, D., Chen, M., Lee, H., Ngiam, J., Le, Q.~V., Wu, Y., et~al.
\newblock Gpipe: Efficient training of giant neural networks using pipeline parallelism.
\newblock \emph{Advances in neural information processing systems}, 32, 2019.

\bibitem[Jangda et~al.(2022)Jangda, Huang, Liu, Sabet, Maleki, Miao, Musuvathi, Mytkowicz, and Saarikivi]{coconet}
Jangda, A., Huang, J., Liu, G., Sabet, A. H.~N., Maleki, S., Miao, Y., Musuvathi, M., Mytkowicz, T., and Saarikivi, O.
\newblock Breaking the computation and communication abstraction barrier in distributed machine learning workloads.
\newblock In \emph{Proceedings of the 27th ACM International Conference on Architectural Support for Programming Languages and Operating Systems}, ASPLOS '22, pp.\  402–416, New York, NY, USA, 2022. Association for Computing Machinery.
\newblock ISBN 9781450392051.
\newblock \doi{10.1145/3503222.3507778}.
\newblock URL \url{https://doi.org/10.1145/3503222.3507778}.

\bibitem[Jiang et~al.(2024{\natexlab{a}})Jiang, Sablayrolles, Roux, Mensch, Savary, Bamford, Chaplot, Casas, Hanna, Bressand, et~al.]{mixtral}
Jiang, A.~Q., Sablayrolles, A., Roux, A., Mensch, A., Savary, B., Bamford, C., Chaplot, D.~S., Casas, D. d.~l., Hanna, E.~B., Bressand, F., et~al.
\newblock {Mixtral} of experts.
\newblock \emph{arXiv preprint arXiv:2401.04088}, 2024{\natexlab{a}}.

\bibitem[Jiang et~al.(2024{\natexlab{b}})Jiang, Tian, Jia, Zheng, Wu, and Wang]{lancet}
Jiang, C., Tian, Y., Jia, Z., Zheng, S., Wu, C., and Wang, Y.
\newblock Lancet: Accelerating mixture-of-experts training via whole graph computation-communication overlapping.
\newblock \emph{Proceedings of Machine Learning and Systems}, 6:\penalty0 74--86, 2024{\natexlab{b}}.

\bibitem[Kaplan et~al.(2020)Kaplan, McCandlish, Henighan, Brown, Chess, Child, Gray, Radford, Wu, and Amodei]{kaplan2020scalinglaws}
Kaplan, J., McCandlish, S., Henighan, T., Brown, T.~B., Chess, B., Child, R., Gray, S., Radford, A., Wu, J., and Amodei, D.
\newblock Scaling laws for neural language models.
\newblock \emph{CoRR}, abs/2001.08361, 2020.
\newblock URL \url{https://arxiv.org/abs/2001.08361}.

\bibitem[Kwon et~al.(2023)Kwon, Li, Zhuang, Sheng, Zheng, Yu, Gonzalez, Zhang, and Stoica]{vllm}
Kwon, W., Li, Z., Zhuang, S., Sheng, Y., Zheng, L., Yu, C.~H., Gonzalez, J.~E., Zhang, H., and Stoica, I.
\newblock Efficient memory management for large language model serving with pagedattention.
\newblock In \emph{Proceedings of the ACM SIGOPS 29th Symposium on Operating Systems Principles}, pp.\  611--626, 2023.

\bibitem[NVIDIA()]{tensorrtllm}
NVIDIA.
\newblock Tensorrt-llm.
\newblock URL \url{https://github.com/NVIDIA/TensorRT-LLM}.

\bibitem[{NVIDIA}(2023)]{MULTIMEM_PTX}
{NVIDIA}.
\newblock {Parallel Thread Execution ISA Version 8.1 -- Cooperative Thread Arrays}.
\newblock \url{https://docs.nvidia.com/cuda/parallel-thread-execution/index.html#changes-in-ptx-isa-version-8-1}, 2023.

\bibitem[{NVIDIA}(2024)]{NVSHARP}
{NVIDIA}.
\newblock {NVIDIA Scalable Hierarchical Aggregation and Reduction Protocol (SHARP)}.
\newblock \url{https://docs.nvidia.com/networking/display/sharpv300}, 2024.

\bibitem[{NVIDIA}(2025)]{NVSHMEM}
{NVIDIA}.
\newblock {NVIDIA OpenSHMEM Library (NVSHMEM)}.
\newblock \url{https://docs.nvidia.com/nvshmem/api/introduction.html}, 2025.

\bibitem[OpenAI(2023)]{openai2022gpt4techreport}
OpenAI.
\newblock {GPT-4} technical report, 2023.
\newblock URL \url{https://doi.org/10.48550/arXiv.2303.08774}.

\bibitem[Patel et~al.(2024)Patel, Choukse, Zhang, Shah, Goiri, Maleki, and Bianchini]{splitwise}
Patel, P., Choukse, E., Zhang, C., Shah, A., Goiri, {\'I}., Maleki, S., and Bianchini, R.
\newblock Splitwise: Efficient generative llm inference using phase splitting.
\newblock In \emph{2024 ACM/IEEE 51st Annual International Symposium on Computer Architecture (ISCA)}, pp.\  118--132. IEEE, 2024.

\bibitem[Prescott(2022)]{nvidia2022setstablepowerstate}
Prescott, R.
\newblock {Advanced API Performance: SetStablePowerState}.
\newblock \url{https://developer.nvidia.com/blog/advanced-api-performance-setstablepowerstate/}, 2022.

\bibitem[{PyTorch Team}(2025)]{pytorch-multimem-allreduce}
{PyTorch Team}.
\newblock Multimem all reduce, 2025.
\newblock URL \url{https://github.com/pytorch/pytorch/blob/v2.6.0/torch/csrc/distributed/c10d/CUDASymmetricMemoryOps.cu}.

\bibitem[{SGLang Team}(2025)]{sglang-tbo}
{SGLang Team}.
\newblock {Deploying DeepSeek with PD Disaggregation and Large-Scale Expert Parallelism on 96 H100 GPUs}.
\newblock \url{https://lmsys.org/blog/2025-05-05-large-scale-ep/}, 2025.

\bibitem[{ShareGPT}(2023)]{sharegpt_vicuna_unfiltered}
{ShareGPT}.
\newblock Sharegpt vicuna unfiltered dataset.
\newblock \url{https://huggingface.co/datasets/anon8231489123/ShareGPT_Vicuna_unfiltered}, 2023.
\newblock Dataset.

\bibitem[Shi et~al.(2023)Shi, Pan, Chu, and Li]{pipemoe}
Shi, S., Pan, X., Chu, X., and Li, B.
\newblock Pipemoe: Accelerating mixture-of-experts through adaptive pipelining.
\newblock In \emph{IEEE INFOCOM 2023-IEEE Conference on Computer Communications}, pp.\  1--10. IEEE, 2023.

\bibitem[Shoeybi et~al.(2019)Shoeybi, Patwary, Puri, LeGresley, Casper, and Catanzaro]{megatron}
Shoeybi, M., Patwary, M., Puri, R., LeGresley, P., Casper, J., and Catanzaro, B.
\newblock Megatron-lm: Training multi-billion parameter language models using model parallelism, 2019.

\bibitem[{TensorFlow Team}(2021)]{xla}
{TensorFlow Team}.
\newblock Xla: Optimizing compiler for tensorflow.
\newblock \url{https://www.tensorflow.org/xla}, 2021.

\bibitem[Vaswani et~al.(2017)Vaswani, Shazeer, Parmar, Uszkoreit, Jones, Gomez, Kaiser, and Polosukhin]{attentionpaper}
Vaswani, A., Shazeer, N., Parmar, N., Uszkoreit, J., Jones, L., Gomez, A.~N., Kaiser, L., and Polosukhin, I.
\newblock Attention is all you need, 2017.

\bibitem[{vLLM Contributors}(2023)]{vllm-rmsnorm-kernel}
{vLLM Contributors}.
\newblock {vLLM: Fused Add and Norm Kernel}.
\newblock \url{https://github.com/vllm-project/vllm/blob/v0.8.5/csrc/layernorm_kernels.cu}, 2023.

\bibitem[{vLLM Team}(2025{\natexlab{a}})]{vllmchunkedprefill}
{vLLM Team}.
\newblock Optimization and tuning.
\newblock \url{https://docs.vllm.ai/en/v0.8.5/performance/optimization.html}, 2025{\natexlab{a}}.

\bibitem[{vLLM Team}(2025{\natexlab{b}})]{vllmv1}
{vLLM Team}.
\newblock vllm v1, 2025{\natexlab{b}}.
\newblock URL \url{https://blog.vllm.ai/2025/01/27/v1-alpha-release.html}.

\bibitem[Wang et~al.(2022)Wang, Wei, Sabne, Davis, Ilbeyi, Hechtman, Chen, Murthy, Maggioni, Zhang, Kumar, Guo, Xu, and Zhou]{googleoverlap}
Wang, S., Wei, J., Sabne, A., Davis, A., Ilbeyi, B., Hechtman, B., Chen, D., Murthy, K.~S., Maggioni, M., Zhang, Q., Kumar, S., Guo, T., Xu, Y., and Zhou, Z.
\newblock Overlap communication with dependent computation via decomposition in large deep learning models.
\newblock In \emph{Proceedings of the 28th ACM International Conference on Architectural Support for Programming Languages and Operating Systems, Volume 1}, ASPLOS 2023, pp.\  93–106, New York, NY, USA, 2022. Association for Computing Machinery.
\newblock ISBN 9781450399159.
\newblock \doi{10.1145/3567955.3567959}.
\newblock URL \url{https://doi.org/10.1145/3567955.3567959}.

\bibitem[Wang et~al.(2024)Wang, He, Wright, Wehrstedt, Liu, and Liang]{async-tp}
Wang, Y., He, H., Wright, L., Wehrstedt, L., Liu, T., and Liang, W.
\newblock Distributed w/ torchtitan: Introducing async tensor parallelism in pytorch, 2024.
\newblock URL \url{https://discuss.pytorch.org/t/distributed-w-torchtitan-introducing-async-tensor-parallelism-in-pytorch/209487}.

\bibitem[Wang et~al.(2025)Wang, He, and Wehrstedt]{pytorchsymmetricmemory}
Wang, Y., He, H., and Wehrstedt, L.
\newblock Pytorch symmetricmemory: Harnessing nvlink programmability with ease, 2025.
\newblock URL \url{https://dev-discuss.pytorch.org/t/pytorch-symmetricmemory-harnessing-nvlink-programmability-with-ease/2798/1}.

\bibitem[Yang et~al.(2024)Yang, Yang, Zhang, Hui, Zheng, Yu, Li, Liu, Huang, Wei, et~al.]{qwen2.5}
Yang, A., Yang, B., Zhang, B., Hui, B., Zheng, B., Yu, B., Li, C., Liu, D., Huang, F., Wei, H., et~al.
\newblock {Qwen2.5} technical report.
\newblock \emph{arXiv preprint arXiv:2412.15115}, 2024.

\bibitem[Zhang et~al.(2025)Zhang, Zheng, Lin, Jiang, Bao, Jiang, Hou, Cui, Zheng, Chang, Chen, and Liu]{comet}
Zhang, S., Zheng, N., Lin, H., Jiang, Z., Bao, W., Jiang, C., Hou, Q., Cui, W., Zheng, S., Chang, L.-W., Chen, Q., and Liu, X.
\newblock Comet: Fine-grained computation-communication overlapping for mixture-of-experts, 2025.

\bibitem[Zheng et~al.(2024)Zheng, Yin, Xie, Sun, Huang, Yu, Cao, Kozyrakis, Stoica, Gonzalez, Barrett, and Sheng]{sglang}
Zheng, L., Yin, L., Xie, Z., Sun, C., Huang, J., Yu, C.~H., Cao, S., Kozyrakis, C., Stoica, I., Gonzalez, J.~E., Barrett, C., and Sheng, Y.
\newblock Sglang: efficient execution of structured language model programs.
\newblock In \emph{Proceedings of the 38th International Conference on Neural Information Processing Systems}, NIPS '24, Red Hook, NY, USA, 2024. Curran Associates Inc.
\newblock ISBN 9798331314385.

\bibitem[Zheng et~al.(2025{\natexlab{a}})Zheng, Bao, Hou, Zheng, Fang, Huang, Li, Duanmu, Chen, Xu, Guo, Zheng, Jiang, Di, Wang, Ye, Lin, Chang, Lu, Liang, Zhai, and Liu]{tritondistributed}
Zheng, S., Bao, W., Hou, Q., Zheng, X., Fang, J., Huang, C., Li, T., Duanmu, H., Chen, R., Xu, R., Guo, Y., Zheng, N., Jiang, Z., Di, X., Wang, D., Ye, J., Lin, H., Chang, L.-W., Lu, L., Liang, Y., Zhai, J., and Liu, X.
\newblock Triton-distributed: Programming overlapping kernels on distributed ai systems with the triton compiler.
\newblock 2025{\natexlab{a}}.
\newblock URL \url{https://arxiv.org/abs/2504.19442}.

\bibitem[Zheng et~al.(2025{\natexlab{b}})Zheng, Fang, Zheng, Hou, Bao, Zheng, Jiang, Wang, Ye, Lin, et~al.]{tilelink}
Zheng, S., Fang, J., Zheng, X., Hou, Q., Bao, W., Zheng, N., Jiang, Z., Wang, D., Ye, J., Lin, H., et~al.
\newblock Tilelink: Generating efficient compute-communication overlapping kernels using tile-centric primitives.
\newblock \emph{arXiv preprint arXiv:2503.20313}, 2025{\natexlab{b}}.

\bibitem[Zhong et~al.(2024)Zhong, Liu, Chen, Hu, Zhu, Liu, Jin, and Zhang]{distserv}
Zhong, Y., Liu, S., Chen, J., Hu, J., Zhu, Y., Liu, X., Jin, X., and Zhang, H.
\newblock $\{$DistServe$\}$: Disaggregating prefill and decoding for goodput-optimized large language model serving.
\newblock In \emph{18th USENIX Symposium on Operating Systems Design and Implementation (OSDI 24)}, pp.\  193--210, 2024.

\bibitem[Zhu et~al.(2024)Zhu, Zhao, Zhao, Zuo, Gu, Xie, Gao, Xu, Tang, Ye, Kamahori, Lin, Wang, Krishnamurthy, and Kasikci]{nanoflow}
Zhu, K., Zhao, Y., Zhao, L., Zuo, G., Gu, Y., Xie, D., Gao, Y., Xu, Q., Tang, T., Ye, Z., Kamahori, K., Lin, C.-Y., Wang, S., Krishnamurthy, A., and Kasikci, B.
\newblock Nanoflow: Towards optimal large language model serving throughput, 2024.

\bibitem[Zou(2025)]{zou2025symmetric_memory_torch_compile}
Zou, R.
\newblock {[RFC] Support symmetric memory in torch.compile}.
\newblock \url{https://github.com/pytorch/pytorch/issues/162859}, 2025.
\newblock PyTorch GitHub issue \#162859.

\end{thebibliography}
\bibliographystyle{mlsys2026}

\clearpage
\appendix
\section{Implementation Details}
\label{appendix:sec:implementation-details}
\subsection{Fused AllReduce--RMSNorm Kernel}
\label{appendix:sec:implementation-details:fused-allreduce-rmsnorm}
\autoref{lst:fused-ar-rmsnorm-pseudo-code} presents the implementation of our fused AllReduce--RMSNorm CUDA kernel. Each CTA processes a subset of tokens, performs inter-GPU reduction using the \texttt{multimem\_ld\_reduce\_add} primitive, computes the local variance, and applies normalization before storing results via \texttt{multimem\_st}. By eliminating intermediate memory accesses and offloading reduction to NVSwitch, this fused implementation achieves lower HBM communication latency and higher throughput compared to separate AllReduce and RMSNorm kernels.

\subsection{Smart-Splitting}
\label{appendix:sec:implementation-details:smart-splitting}
\Smartsplitting targets the total number of waves in the two split kernels to be not more than those in the original kernel. To achieve this, instead of assigning an equal number of tokens to each split, we increase the number of tokens in the prefix split by the \texttt{split\_offset} to ensure that it has full SM occupancy in its last wave of computation. Note that for very small batches where the original unsplit kernel has only one total wave, this assigns all tokens to the prefix split, effectively disabling splitting. This is discussed further in \S\ref{sec:selective-splitting}.

Since CTA assignment to SMs is a deterministic function of matmul shape, kernel implementation, configuration parameters (e.g. tile size), etc., we can determine the \texttt{split\_offset} either analytically or via offline profiling. However, we use closed-source cuBLAS kernels for matrix multiplication because of their higher performance, making it difficult to determine this split analytically. Thus, we do a simple profiling sweep for a range of sequence lengths, as shown in \autoref{alg:smart_splitting_heuristics}, to determine the best \texttt{split\_offset} for each configuration and use that during runtime.

\begin{figure}[t!]
\centering
\begin{code}
template <typename scalar_t, int width>
__global__ multimem_fused_allreduce_rmsnorm_kernel(...) {
  const int vec_hidden_size = hidden_size / width;
  int tokens_per_cta = (num_tokens + gridDim.x - 1) / gridDim.x;

  sync_remote_blocks<MemOpSem::Relaxed>(signal_pads, rank, world_size);
  __syncthreads();

  for (int iter = 0; iter < tokens_per_cta; iter++) {
    int token_id = blockIdx.x + iter * gridDim.x;
    if (token_id >= num_tokens) continue;

    float variance[1] = {0.0f};
    __shared__ float s_variance;
    int offset = token_id * vec_hidden_size; 
    int offset_scalar = token_id * hidden_size;
    auto input_o = input_v + offset; 
    auto residual_o = residual_v + offset;

    for (int idx = threadIdx.x; idx < vec_hidden_size; idx += blockDim.x) {
      auto multimem_temp = multimem_ld_reduce_add<16>(multimem_address_ptr +                                   
                    offset_scalar + idx * width);
      vec_t temp = *(reinterpret_cast<vec_t*>(&multimem_temp));
      temp += residual_o[idx];
      variance[0] += temp.sum_squares();
      residual_o[idx] = temp;
    }

    blockReduceSum<float, 1>(variance);
    if (threadIdx.x == 0) 
        s_variance = rsqrtf(variance[0] / hidden_size + epsilon);
    __syncthreads();

    for (int idx = threadIdx.x; idx < vec_hidden_size; idx += blockDim.x) {
      vec_t temp = residual_o[idx] * s_variance * weight_v[idx];
      multimem_st<16>(mcptr + offset + idx * width, 
                        *(reinterpret_cast<Vec<16>*>(&temp)));
    }
  }
  __syncthreads();
  sync_remote_blocks<MemOpSem::AcqRel>(signal_pads, rank, world_size);
}
\end{code}
\caption{Implementation of fused AllReduce--RMSNorm kernel. The current implementation only supports BFloat16. For brevity, we omit non-essential portions of the code. The kernel builds upon the PyTorch Multimem AllReduce~\cite{pytorch-multimem-allreduce} and the RMSNorm kernel from vLLM~\cite{vllm-rmsnorm-kernel}.}
\label{lst:fused-ar-rmsnorm-pseudo-code}
\end{figure}

\begin{algorithm}[!tb]
\begin{algorithmic}[1]
\REQUIRE A set of batch sizes $B_{\text{set}}$, a set of sequence lengths $L_{\text{set}}$, token limit $\textit{MAX\_TOKENS}$
\STATE Initialize empty table $\text{optimal\_split}[B,L]$
\FOR{each batch size $B$ in $B_{\text{set}}$}
    \FOR{each sequence length $L$ in $L_{\text{set}}$}
        \STATE $\text{num\_tokens} \gets B \times L$
        \IF{$\text{num\_tokens} > \textit{MAX\_TOKENS}$}
            \STATE \textbf{continue}
        \ENDIF
        \STATE half\_tokens $\gets \text{num\_tokens} / 2$
        \STATE best\_offset $\gets 0$
        \STATE min\_time $\gets \infty$
        \FOR{offset in $\{0, 64, 128, 192, 256, 512\}$}
            \IF{offset $\geq$ half\_tokens}
                \STATE \textbf{continue}
            \ENDIF
            \STATE $\text{time} \gets \text{model.forward}(B, L, \text{half\_tokens} + \text{offset}, \text{half\_tokens} - \text{offset})$
            \IF{time $<$ min\_time}
                \STATE min\_time $\gets$ time
                \STATE best\_offset $\gets$ offset
            \ENDIF
        \ENDFOR
        \STATE $\text{optimal\_split}[B,L] \gets \text{best\_offset}$
    \ENDFOR
\ENDFOR
\end{algorithmic}
\caption{\Smartsplitting}
\label{alg:smart_splitting_heuristics}
\end{algorithm}

\subsection{Selective Enabling of Splitting and Overlap}
\label{sec:selective-splitting}

As discussed earlier, splitting a batch into smaller sub-batches can result in overheads in both compute and communication (e.g., see \autoref{fig:reducescatter-vs-size}). \Smartsplitting and overlap mitigate most of these overheads and provide significant gains even at smaller batches of 1K tokens (\autoref{fig:end-to-end-perf-prefill-latency-8h100}). However, if the batch has very few tokens, these splitting overheads can be significant and eat into the gains from overlapping compute and communication. In such scenarios, we disable the splitting and overlap, but continue to use our fused AllReduce--RMSNorm kernel. This is achieved by a simple thresholding as shown in \autoref{fig:sysname.drawio}. Based on offline profiling, we use thresholds of $1K$ for the dense Llama and Qwen models, and $4K$ for the Mixtral MoE model on $8\times$H100 DGX system.

\section{Evaluation}
\label{appendix:sec:evaluation}
We provide some additional evaluations of \sysname in this section.

\subsection{Throughput Gains}
\label{appendix:sec:evaluation:throughput}
\autoref{fig:sysname-end-to-end-perf-hybrid-4gpus} presents \sysname's throughput gains on $4\times$H100 GPUs for various end-to-end workload traces. Similar to the 8-GPU results discussed in the main paper, \sysname consistently improves throughput across both real traces such as ShareGPT and arXiv, as well as synthetic traces with fixed input and output lengths. While the relative performance is lesser as compared to 8 GPUs because of lower communication overheads, the trends remain consistent, demonstrating that \sysname's optimizations scale effectively even at smaller GPU counts.
\begin{figure}[tb]
  \centering
  \includegraphics[width=\linewidth]{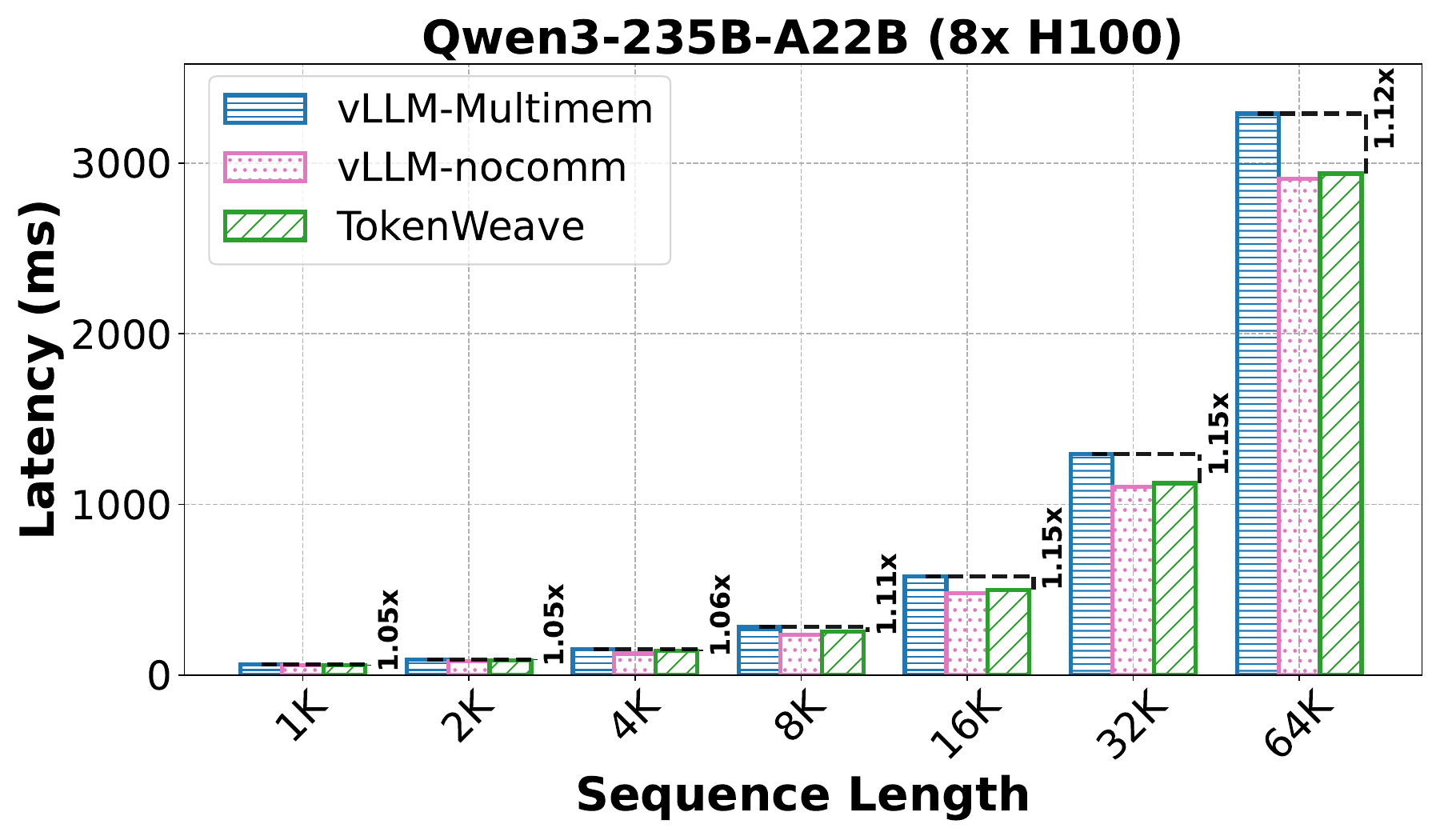}
  \vspace{-22pt}
  \caption{Shown are the execution times for prefill requests with varying sequence lengths for Qwen3-235B-A22B on an $8\times$H100 DGX system.}
\label{fig:sysname-end-to-end-perf-prefill-qwen3moe-8GPUs}
\end{figure}

\subsection{Latency Gains}
\label{appendix:sec:evaluation:latency}
\begin{figure*}[!tb]
  \centering
    \includegraphics[width=0.75\textwidth]{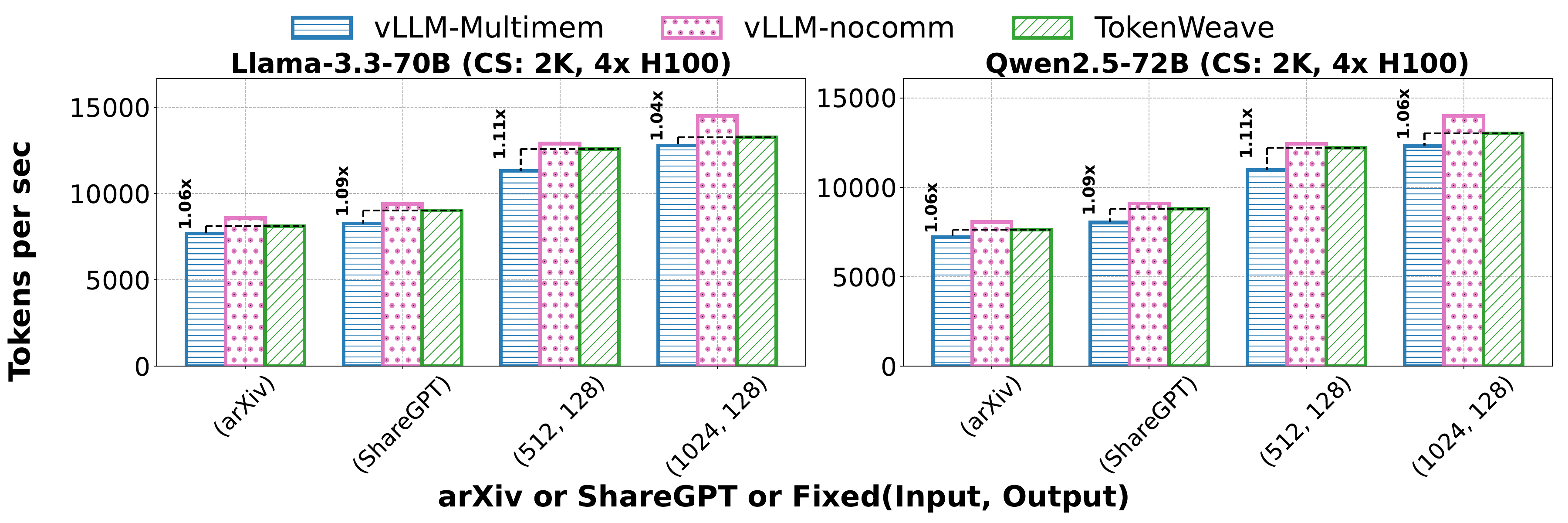}
    \vspace{-12pt}
    \caption{{\bf \sysname throughput gains for end-to-end workload traces.} Shown are throughput measurements across fixed (input, output)-length traces, as well as ShareGPT and arXiv traces, for two models on $4\times$H100.}
\label{fig:sysname-end-to-end-perf-hybrid-4gpus}
\end{figure*}
\begin{figure*}[!tb]
  \centering
    \includegraphics[width=0.8\textwidth]{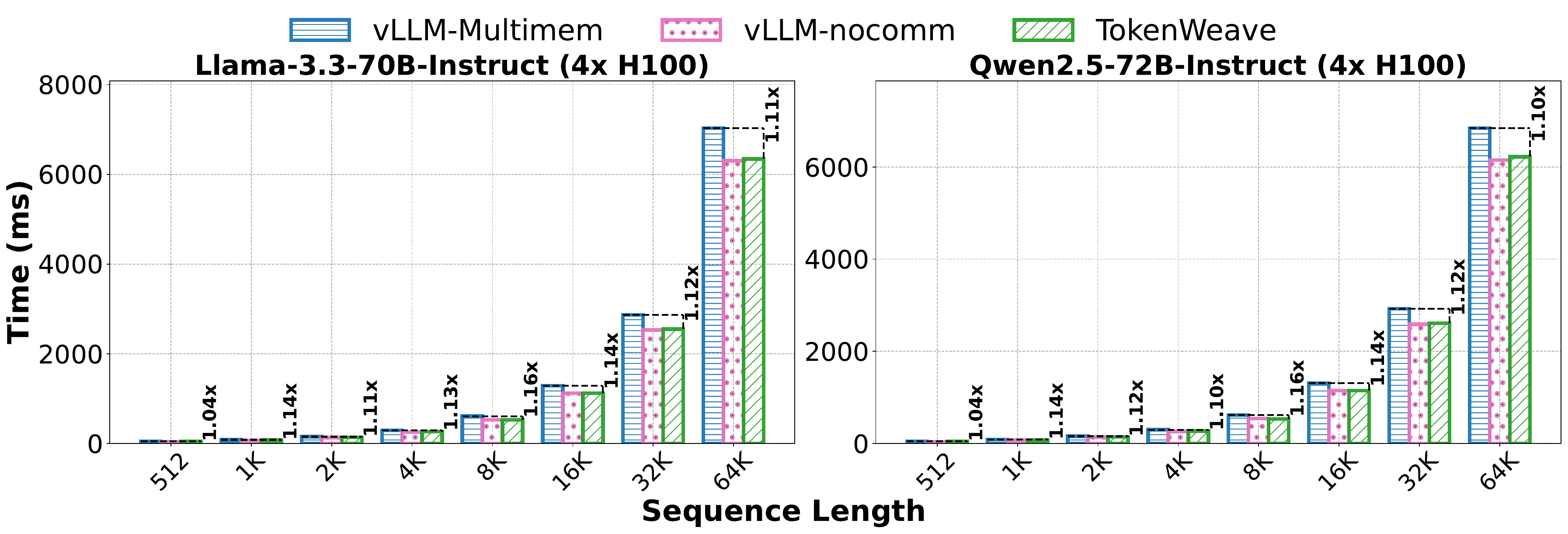}
    \vspace{-12pt}
    \caption{{\bf \sysname latency gains.} Shown are the execution times for prefill requests with varying sequence lengths for different models on $4\times$H100.}
\label{fig:sysname-end-to-end-perf-prefill-4GPUs}
\end{figure*}

\noindent\textbf{Qwen/Qwen3-235B-A22B on $8\times$H100:} \autoref{fig:sysname-end-to-end-perf-prefill-qwen3moe-8GPUs} provides the latency improvements of \sysname on $8\times$H100 GPUs for Qwen3-235B-A22B with varying sequence lengths. As shown, the gains are in a similar range as for the Mixtral MoE model shown in \autoref{fig:end-to-end-perf-prefill-latency-8h100}.

\noindent\textbf{Varying sequence lengths on $4\times$H100:} \autoref{fig:sysname-end-to-end-perf-prefill-4GPUs} provides the latency improvements of \sysname on $4\times$H100 GPUs for various models and varying sequence lengths. Similar to the 8-GPU results discussed in the main paper, \sysname provides substantial gains and is, in fact, close to or better than the theoretical \textit{vLLM-nocomm} baseline with zero communication overhead. Thus, \sysname not only recovers all communication overhead but also provides additional gains due to RMSNorm fusion.

\begin{figure*}[!tb]
  \centering
  \begin{subfigure}{\textwidth}
    \centering
    \includegraphics[width=0.99\textwidth]{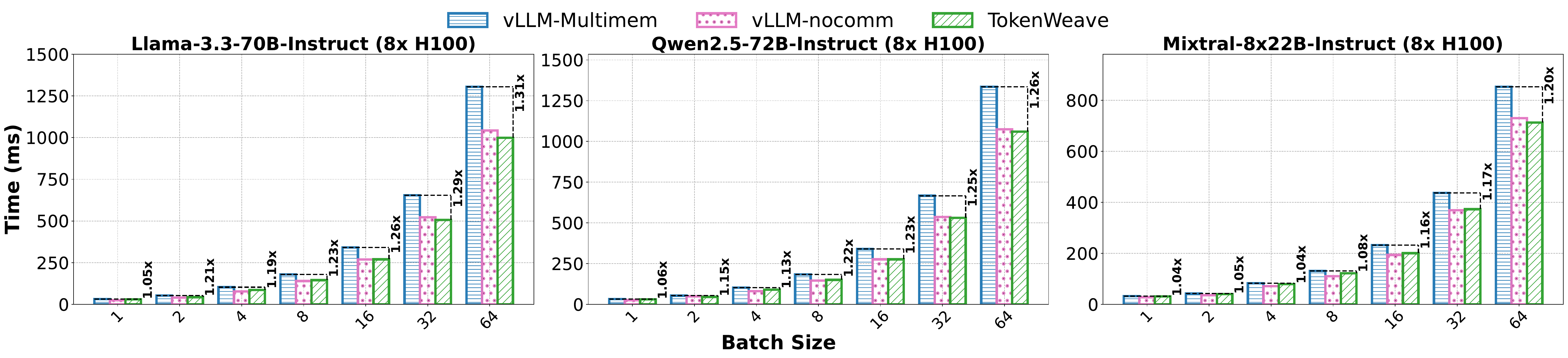}
    \caption{$8\times$H100}
    \label{fig:8xh100-seq512-vary-batch}
  \end{subfigure}

  \begin{subfigure}{\textwidth}
    \centering
    \includegraphics[width=0.8\textwidth]{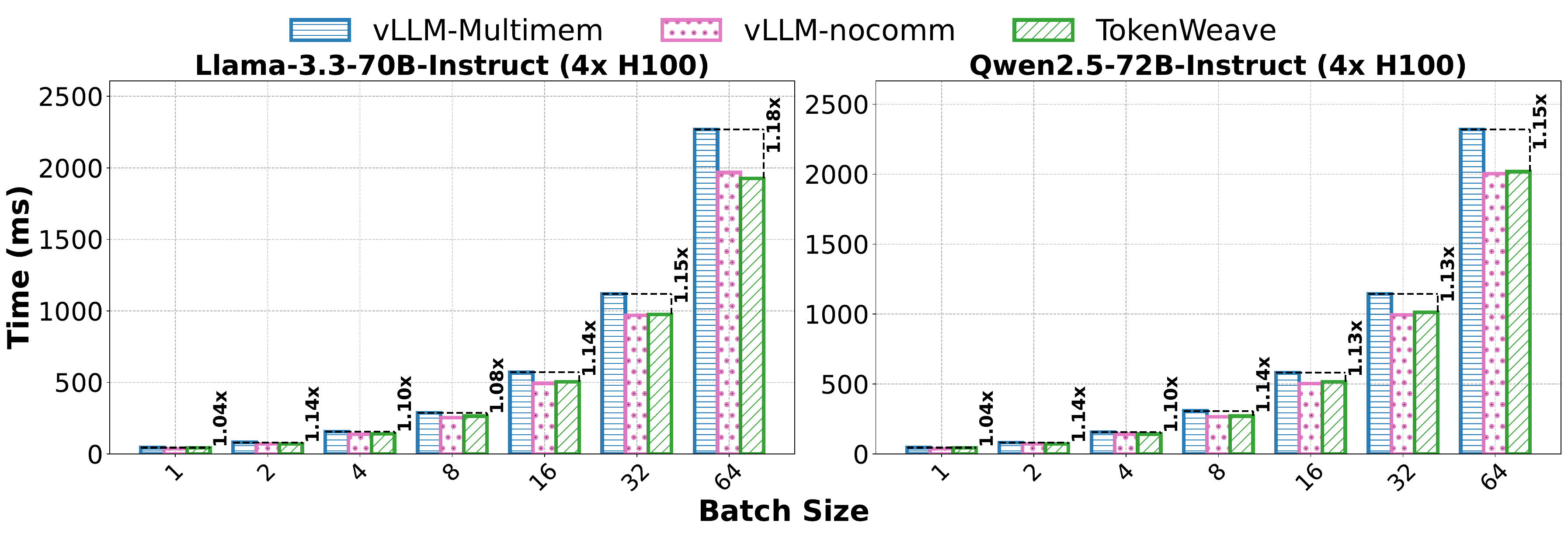}
    \caption{$4\times$H100}
    \label{fig:4xh100-seq512-vary-batch}
  \end{subfigure}
  \vspace{-18pt}
  \caption{{\bf \sysname latency gains.} Shown are the execution times for prefill requests with varying batch sizes and sequence length of 512 for different models on (a) $8\times$H100 and (b) $4\times$H100. In almost all cases, \sysname is close to or better than the theoretical \textit{vLLM-nocomm} baseline with zero communication overhead, showing that \sysname not only recovers all communication overhead but also provides additional gains due to RMSNorm fusion.}
\label{fig:sysname-end-to-end-perf-prefill-varying-batch}
\end{figure*}
\label{sec:latency-gains-varying-batch-size}
\noindent\textbf{Varying batch sizes:} The splitting in \sysname is performed at the token-level and works with any batch size. Similarly, the performance is also a function of mainly the total number of tokens in the batch and does not depend on the batch size directly. However, for completeness, we also show results with varying batch sizes in \autoref{fig:sysname-end-to-end-perf-prefill-varying-batch}. As shown, the results stay similar to the case of fixed batch size and varying sequence lengths (\autoref{fig:end-to-end-perf-prefill-latency-8h100} and \autoref{fig:sysname-end-to-end-perf-prefill-4GPUs}).

\subsection{Ablation Study}
\label{appendix:sec:evaluation:ablation}
\begin{figure*}
  \centering
    \includegraphics[width=0.8\textwidth]{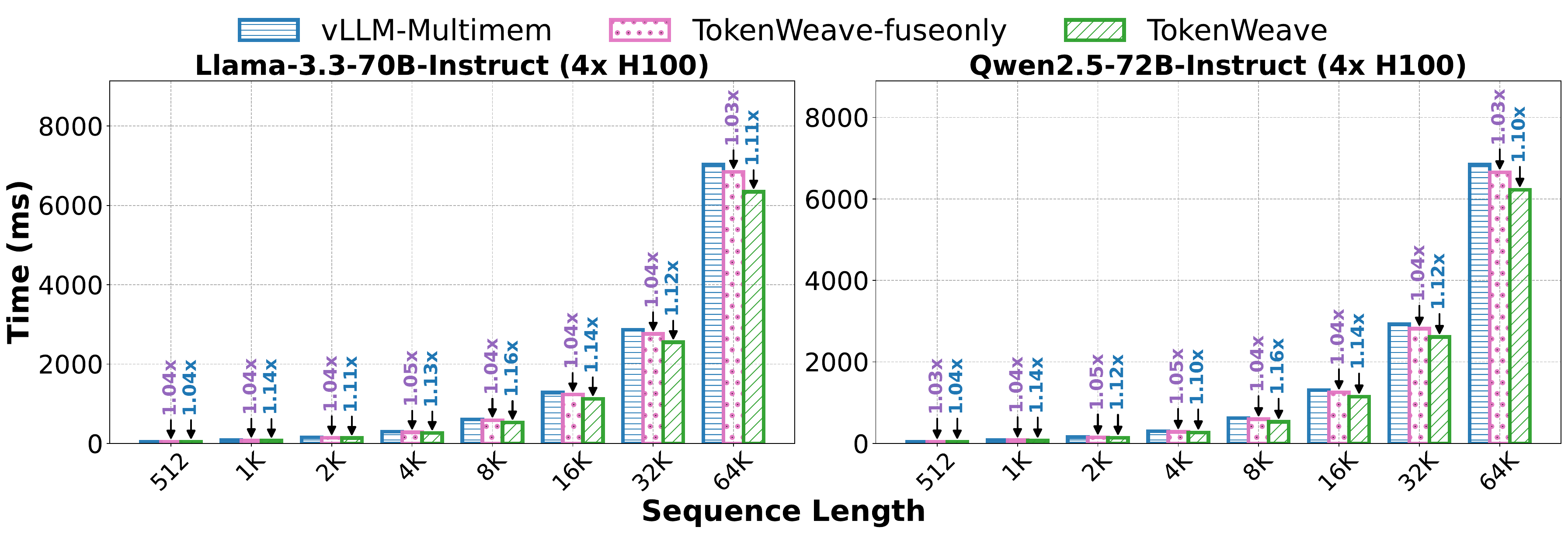}
    \caption{{\bf \sysname Fused AllReduce--RMSNorm kernel ablation.} We compare \textit{\sysname-fuseonly} and full \sysname against the \textit{vLLM-Multimem} baseline. Execution times are shown for prefill requests with varying sequence lengths for different models on $4\times$H100. \textit{\sysname-fuseonly} provides gains due to the elimination of redundancy in RMSNorm computation and intermediate memory accesses, while \sysname provides additional gains from the compute-communication overlap.}
  \label{fig:sysname-end-to-end-perf-prefill-fuse-only-4h100}
\end{figure*}

\autoref{fig:sysname-end-to-end-perf-prefill-fuse-only-4h100} shows ablation results for the fused AllReduce--RMSNorm kernel for the 4-GPU configuration, in addition to the 8-GPU results discussed in the main paper. The fused kernel remains effective even on 4 GPUs, although the relative performance improvement is smaller compared to 8 GPUs. This is because the compute savings in RMSNorm from the reordering after ReduceScatter are $\mathcal{O}(N)$, where $N$ is the number of GPUs.

\begin{figure*}[!tb]
  \centering
  \begin{subfigure}{\textwidth}
    \centering
    \includegraphics[width=0.99\textwidth]{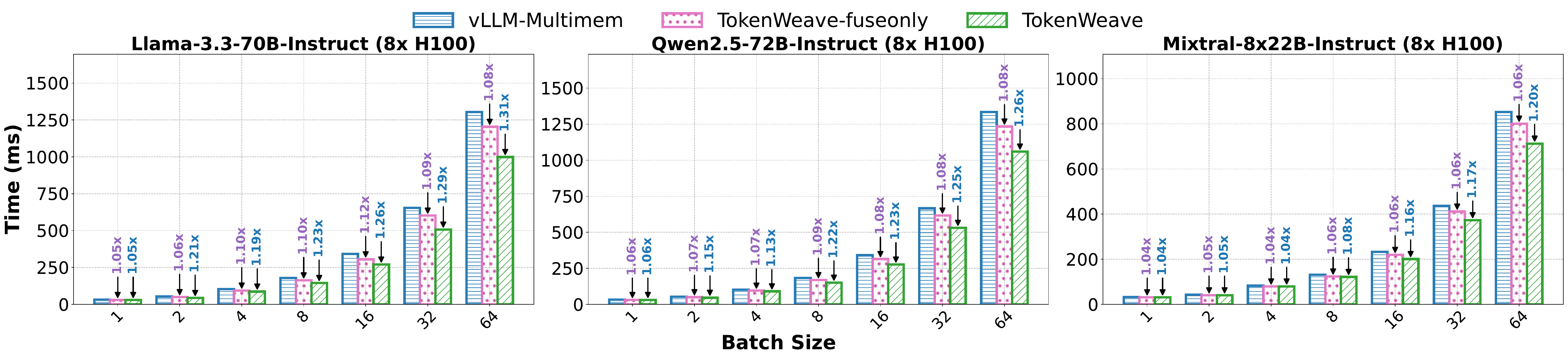}
    \caption{$8\times$H100}
    \label{fig:prefill_fuseonly_8xh100_batchsize}
  \end{subfigure}
    \begin{subfigure}{\textwidth}
    \centering
    \includegraphics[width=0.8\textwidth]{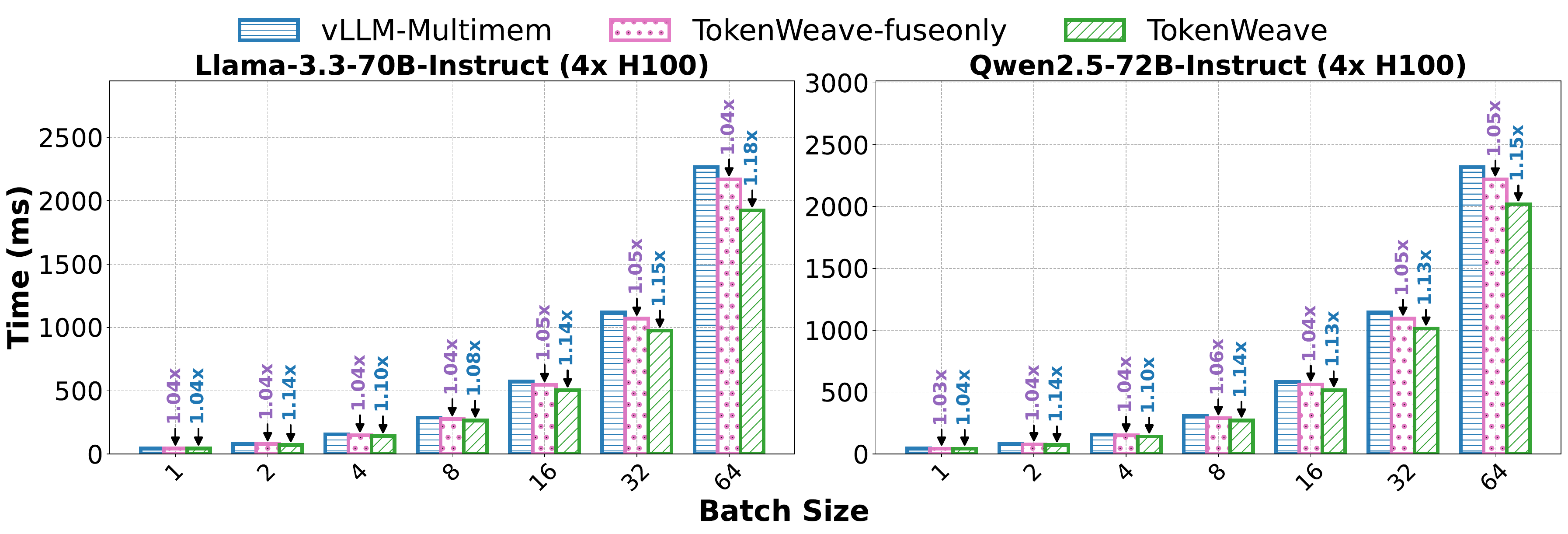}
    \caption{$4\times$H100}
    \label{fig:prefill_fuseonly_4xh100_batch_size}
  \end{subfigure}
  \caption{{\bf \sysname Fused AllReduce--RMSNorm kernel ablation.} Shown are the execution times of prefill requests with varying batch sizes, with the sequence length fixed at 512, for different models on (a) $8\times$H100 and (b) $4\times$H100. \textit{\sysname-fuseonly} provides gains due to the elimination of redundancy in RMSNorm computation and intermediate memory accesses, while \sysname provides additional gains through compute-communication overlap.}
\label{fig:sysname-end-to-end-perf-prefill-fuse-only-varying-batch}
\end{figure*}

For completeness, we also show the ablation results across varying batch sizes in \autoref{fig:sysname-end-to-end-perf-prefill-fuse-only-varying-batch}. As expected, the results stay similar to the case of fixed batch size and varying sequence lengths (\autoref{fig:sysname-end-to-end-perf-prefill-fuse-only} and \autoref{fig:sysname-end-to-end-perf-prefill-fuse-only-4h100}).

\section{Evaluation on NVIDIA DGX B200 System}
\label{appendix:sec:b200:eval}
\begin{table*}[t]
\centering
\small
\setlength{\tabcolsep}{6pt}
\renewcommand{\arraystretch}{2.1} 
\resizebox{\linewidth}{!}{%
\begin{tabular}{lcccccccccccc}
\toprule
\rowcolor{headercolor}
& \multicolumn{4}{|c|}{Low} & \multicolumn{4}{|c|}{Mid} & \multicolumn{4}{|c|}{High} \\
\toprule
\rowcolor{headercolor}
\# Tokens & 32 & 64 & 128 & 256 & 512 & 1K & 2K & 4K & 8K & 16K & 32K & 64K \\

\midrule

\rowcolor{lowcolor}
AR 
& 26.08 & 28.80 & 32.29 & 35.20 & 45.55 & 60.26 & 95.86 & 166.61 & 305.78 & 578.48 & 1131.55 & 2240.93 \\

\rowcolor{lowcolor}
RMSNorm
& 14.46 & 13.15 & 14.67 & 13.66 & 15.38 & 21.12 & 31.62 & 53.66 & 93.38 & 173.84 & 333.02 & 654.51 \\

\hdashline

\rowcolor{midcolor}
AR+RMSNorm 
& 40.54 & 41.95 & 46.96 & 48.86 & 60.93 & 81.38 & 127.47 & 220.27 & 399.15 & 752.32 & 1464.58 & 2895.44 \\

\hdashline

\rowcolor{highcolor}
Fused (Ours) 
& 30.46 & 32.45 & 34.14 & 39.18 & 49.31 & 63.62 & 100.48 & 170.14 & 307.71 & 581.55 & 1130.69 & 2236.02 \\

\hdashline

\rowcolor{lowcolor}
\textbf{Speedup}
& $\mathbf{1.33}\times$ & $\mathbf{1.29}\times$ & $\mathbf{1.38}\times$ & $\mathbf{1.25}\times$ & $\mathbf{1.24}\times$
& $\mathbf{1.28}\times$ & $\mathbf{1.27}\times$ & $\mathbf{1.29}\times$ & $\mathbf{1.30}\times$ & $\mathbf{1.29}\times$
& $\mathbf{1.30}\times$ & $\mathbf{1.29}\times$ \\

\bottomrule
\end{tabular}%
}

\caption{{\bf Fused AllReduce--RMSNorm kernel performance.} Multimem-based AllReduce and RMSNorm latencies are measured independently in isolation. AR+RMSNorm denotes their summed latency (i.e., AllReduce + RMSNorm), which is compared against our fused kernel latency (hidden size $8192$, $us$, $bf16$, $8\times$B200 DGX). Speedup is the ratio of AR+RMSNorm to the fused latency.}
\label{fig:sysname-8xb200-hiddensize-8k-fused-kernel}
\end{table*}

\begin{figure*}[t]
  \centering
    \includegraphics[width=0.8\textwidth]{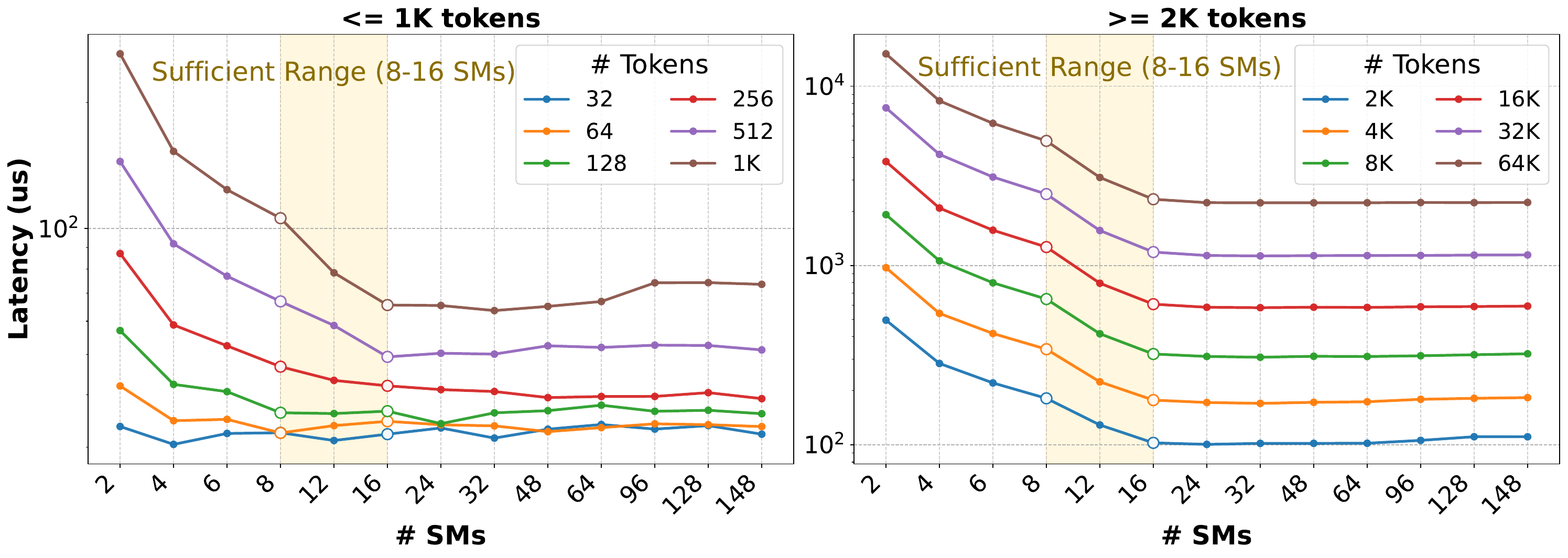}
    \caption{Latency of the fused AllReduce--RMSNorm kernel versus SM count on an $8\times$B200 DGX system. Results are shown for sequence lengths from $32$ to $64K$ tokens (hidden size $8192$, $bf16$). Similar to the $8\times$H100 results, latency reductions diminish beyond roughly $8$--$16$ SMs.}
  \label{fig:sysname-8xb200-hiddensize-8k-cta-scaling}
\end{figure*}
\noindent\textbf{Hardware and Environment:} We run experiments on a server with eight NVIDIA B200 GPUs, each with 192~GB of HBM. The host system has 106 CPU cores and approximately 2.4~TB of system memory. We use \texttt{PyTorch}~2.10.0 with \texttt{CUDA}~13.0. For end-to-end evaluation, we use \texttt{vLLM}~0.14.1 and build \sysname on top of it, since \texttt{vLLM}~0.8.5 does not support B200 GPUs. We use the FlashInfer~0.5.3 backend for attention computation. Unlike our H100 experiments, we do not lock GPU clocks to the TDP frequency, as it requires root access that is unavailable on our shared B200 server. For microbenchmarks, we perform a warm-up phase, flush the L2 cache, and use \verb!cudaEvent!s for timing measurements. For end-to-end (e2e) results, we also perform warm-up and then measure latency using \verb!time.perf_counter()!. We disable prefix caching and ignore detokenization time, consistent with the H100 experiments.

\subsection{Fused AllReduce--RMSNorm Kernel}
To further demonstrate the adaptability of our fused kernel, we evaluate it on an NVIDIA DGX B200 system. This DGX system also provides NVSHARP capabilities and supports Multimem instructions. As depicted in \autoref{fig:sysname-8xb200-hiddensize-8k-fused-kernel}, the fused kernel reduces latency across sequence lengths and achieves $1.24\text{--}1.38\times$ speedup over an additive baseline formed by standalone Multimem AllReduce and RMSNorm. We measure the latency of both Multimem AllReduce and RMSNorm separately in isolation.

Similar to the trend observed in \autoref{fig:allreduce-layernorm-fused-by-SM}, the latency of our fused kernel on $8\times$B200 decreases as the number of allocated SMs increases, with minor improvements beyond a threshold. As shown in \autoref{fig:sysname-8xb200-hiddensize-8k-cta-scaling}, for a hidden size of $8192$ in $bf16$, performance improvements are small beyond approximately $8$--$16$ SMs.

\subsection{Decode-Only Batches}
\label{sec:b200-decode-only}
\begin{figure*}[t]
  \centering
   \includegraphics[width=0.7\textwidth]{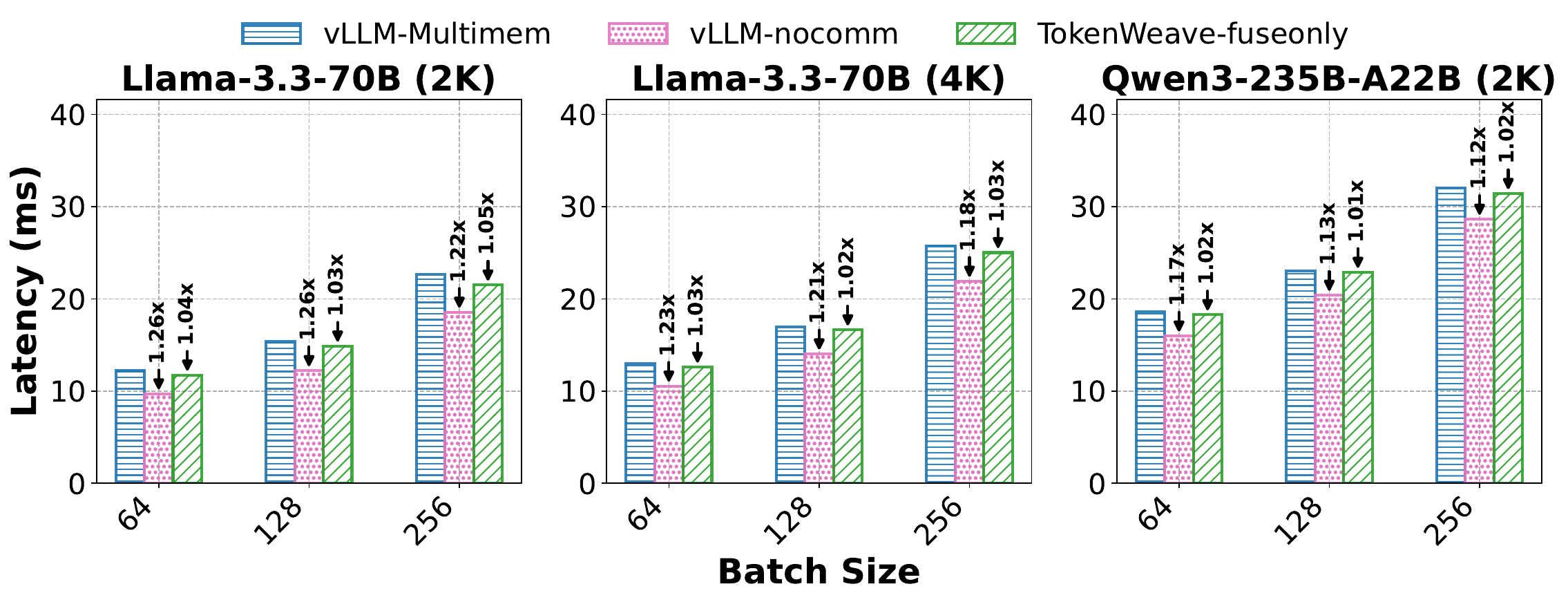}
   \vspace{-12pt}
  \caption{{\bf \sysname Decode Latency Gains on an $8\times$B200} DGX system for Llama-3.3-70B (context lengths $2K$ and $4K$) and Qwen3-235B-A22B (context length $2K$).}
  \label{fig:sysname-decodeonly-8xb200-perf}
\end{figure*}

In the V1 architecture, vLLM uses \texttt{torch.compile} to reduce Python execution overhead during inference. However, at present, \texttt{torch.compile} does not support symmetric memory allocation~\cite{zou2025symmetric_memory_torch_compile}. To evaluate performance at small decode batch sizes, we disable \texttt{torch.compile} in \texttt{vLLM}~0.14.1 but use CUDA Graphs for both the baseline and \sysname configurations. Under small decode batches, the overhead introduced by splitting tokens into two parts offsets the performance gains obtained from compute-communication overlap. Therefore, we report only \textit{\sysname-fuseonly} results for Llama-3.3-70B and Qwen3-235B-A22B on an $8\times$B200 DGX system.

As shown in \autoref{fig:sysname-decodeonly-8xb200-perf}, we see latency improvements of $1.01$--$1.05\times$ across different models and context lengths. Note that Llama-3.3-70B has a hidden dimension of $8$K, whereas Qwen3-235B-A22B has a hidden dimension of $4$K. Thus, the latency gains from fusion alone are smaller for the Qwen3 MoE model than for Llama, as expected. Additionally, increasing the context length reduces the RMSNorm cost as a fraction of the total decode latency, which further decreases the relative performance gains. This trend can be seen in the $2$K vs.\ $4$K results in \autoref{fig:sysname-decodeonly-8xb200-perf}.

\subsection{Prefill-Only Batches}
As mentioned in \S\ref{sec:b200-decode-only}, \texttt{torch.compile} currently does not support symmetric memory. Further, our full solution, which involves wave-aware token splitting and overlapping the compute of one split with the fused AllReduce--RMSNorm of the other, does not support CUDA Graphs. We therefore run prefill latency experiments in eager mode, which disables both \texttt{torch.compile} and CUDA graphs.

\begin{figure*}[t]
    \centering
    \includegraphics[width=0.7\textwidth]{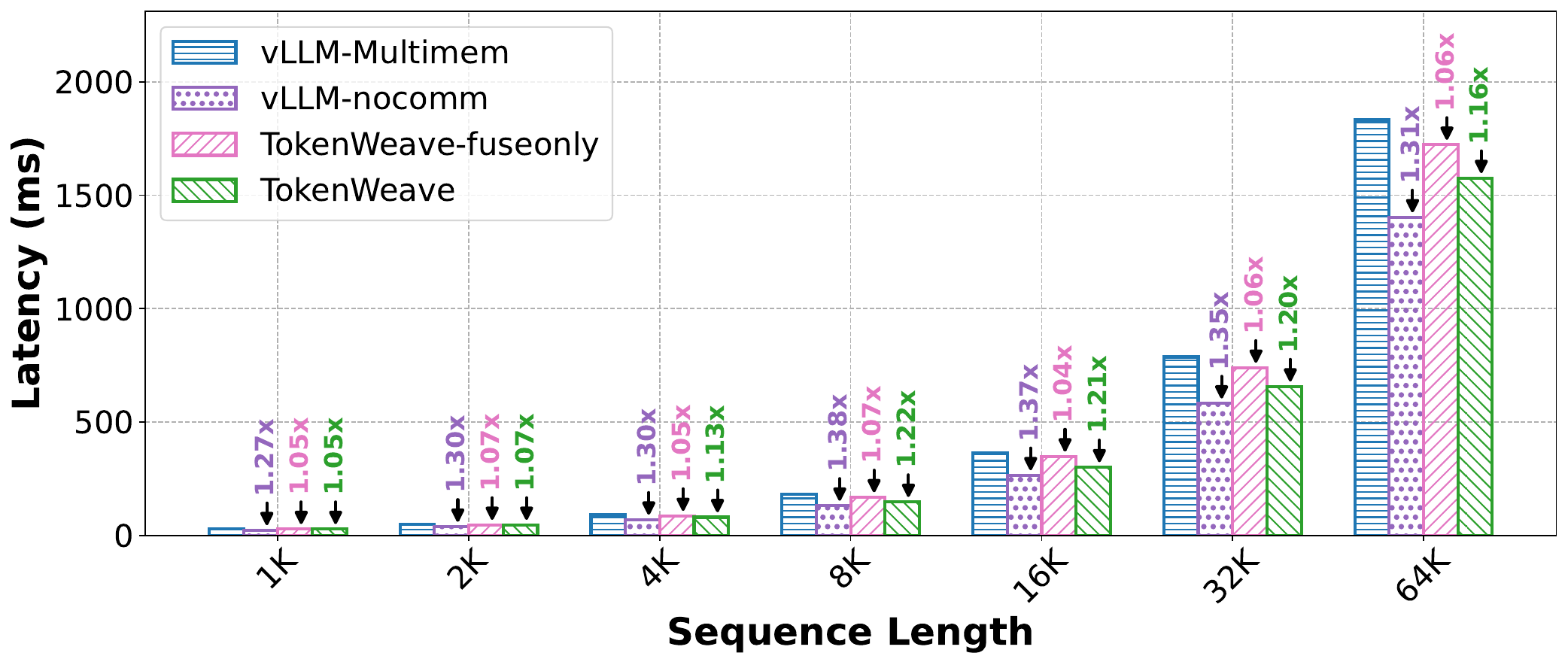}
    \vspace{-6pt}
    \caption{\textbf{\sysname Prefill latency gains on an $8\times$B200} DGX system for Llama-3.3-70B (varying sequence lengths). \textit{\sysname-fuseonly} achieves $1.05\times$--$1.07\times$ speedup, while full \sysname achieves up to $1.22\times$ over \textit{vLLM-Multimem}.}
\vspace{-6pt}
\label{fig:sysname-prefillonly-8xb200-perf}
\end{figure*}

Consistent with H100 results, our fused kernel alone achieves $1.05\times$--$1.07\times$ latency improvements on Llama-3.3-70B. Our full solution, which comprises the fused AllReduce--RMSNorm kernel, smart splitting, and overlap, achieves up to $1.22\times$ speedup over \textit{vLLM-Multimem} across a range of $1K$--$64K$ tokens, as shown in \autoref{fig:sysname-prefillonly-8xb200-perf}.
\section{Artifact}
\label{appendix:sec:artifact}
\subsection{Abstract}

Distributed inference of LLMs can incur overheads of up to 20\%, even when GPUs are connected via high-speed interconnects such as NVLink. Additionally, RMSNorm and residual addition also introduce non-trivial overhead, as they lie on the critical path of execution. \sysname addresses these inefficiencies by proposing (1) a fused kernel that optimizes RMSNorm and residual addition, and (2) a coarse-grained, GPU-wave-aware compute-communication overlap mechanism. 

\subsection{Artifact check-list (meta-information)}

{\small
    \begin{itemize}
      \item {\bf Compilation: }CUDA 12.4
      \item {\bf Model: }Llama-3.3-70B, Qwen2.5-72B, Mixtral-8x22B
      \item {\bf Dataset: }Synthetic and ShareGPT
      \item {\bf Environment: }Python~3.12, PyTorch~v2.6.0, Ubuntu~22.04
      \item {\bf Hardware: }$8\times$H100 DGX with NVSHARP support
      \item {\bf Metrics: }Latency and Tokens per second
      \item {\bf Experiments: }Microbenchmarks and end-to-end results
      \item {\bf How much time is needed to prepare workflow (approximately)?: }30 minutes
      \item {\bf How much time is needed to complete experiments (approximately)?: }~9 hours 25 minutes
      \item {\bf Publicly available?: }Yes
      \item {\bf GitHub: }\url{https://github.com/microsoft/tokenweave}
      \item {\bf DOI: }\url{https://doi.org/10.5281/zenodo.18844243}
    \end{itemize}
}

\subsection{Description}

\subsubsection{How to access}
The source code is available at \url{https://github.com/microsoft/tokenweave} on the \texttt{artifact-evaluation} branch. A snapshot of the code is also archived via DOI: \url{https://doi.org/10.5281/zenodo.18844243}. We recommend using the GitHub repository for the most up-to-date and actively maintained version. Note that root access\footnote{This requirement can be relaxed by commenting out the GPU clock-setting code. However, performance may vary slightly from the reported H100 results if GPU clocks are not fixed to the TDP frequency, but overall trends should remain the same.} is recommended for running this artifact, as the evaluation scripts automatically fix GPU clocks to their TDP frequency.

\subsubsection{Hardware dependencies}
This artifact requires an x86 machine with an NVIDIA $8\times$H100 DGX system, with each H100 GPU having 80 GB of memory.

\subsubsection{Software dependencies}
The artifact has been tested on a machine running Ubuntu~22.04 and CUDA~12.4 with PyTorch~v2.6.0. PyTorch~v2.6.0 provides symmetric memory support. All other dependencies are resolved during installation. 

\subsubsection{Datasets}
Some experiments use the \verb!ShareGPT! dataset, which is downloaded automatically on the fly. For most experiments, we use synthetic random datasets. Note that the arXiv dataset evaluation from the paper is not included in the artifact scripts to reduce overall evaluation time.

\subsubsection{Models}
This artifact evaluates Llama-3.3-70B, Qwen2.5-72B, and Mixtral-8x22B. Note that we use instruction-tuned variants for all evaluated models. Accessing Qwen2.5-72B is straightforward, but accessing Llama-3.3-70B and Mixtral-8x22B requires logging into Hugging Face with your Hugging Face token:
\begin{lstlisting}[style=bash]
huggingface-cli login --token HF_TOKEN
\end{lstlisting}

\subsection{Installation}
To ease the setup, we recommend using one of the following Docker images: \textcolor{teal}{\textit{pytorch/pytorch:2.6.0-cuda12.4-cudnn9-devel}} or \textcolor{teal}{\textit{vllm/vllm-openai:v0.8.5}}.
\subsubsection{Docker Setup}
After launching a container from one of the above images, follow the steps below to set up \sysname.
\begin{lstlisting}[style=bash]
apt-get update; apt-get upgrade -y; apt-get install kmod git build-essential tmux -y   
git clone https://github.com/microsoft/tokenweave.git --single-branch -b artifact-evaluation # Clone the repository
cd tokenweave
# Install miniconda; skip if already installed
make install_miniconda # 30 seconds
make create_env
bash # Refresh shell and activate
conda activate tokenweave
make install # or pip3 install -v -e . (18 minutes)
make install_dependencies # 17 seconds
huggingface-cli login --token HF_TOKEN
\end{lstlisting}
\subsubsection{Run offline inference examples}
To verify that the setup is working correctly, run the following commands.


\begin{lstlisting}[style=bash]
make run_qwen2
make run_mixtral
make run_llama3
\end{lstlisting}
\textbf{Note:} If the Llama 3 download stalls, terminate and restart the process. This typically resolves the issue.
\subsection{Experiment workflow}
The \sysname evaluation scripts are available in the \texttt{artifact/} folder. Our evaluation consists of two types of experiments: microbenchmarks (Table~1; Figures~1, 4, 6, 9, and 10) and end-to-end LLM performance (Figures~2, 11, 12, and 13). Use the Makefile present in the \texttt{artifact/} folder to run experiments as follows:
\begin{lstlisting}[style=bash]
cd artifact
make figure_4_6 # 20 minutes
make table_1_figure_10 # 1 hour 25 minutes
make figure_9 # 8 minutes
make figure_1 # 3 hours 25 minutes
make figure_2_13 # 42 minutes
make figure_11 # 1 hour 32 minutes
make figure_12 # 1 hour 52 minutes
\end{lstlisting}

Alternatively, run the following to execute all experiments and generate all plots in a single step:
\begin{lstlisting}[style=bash]
cd artifact
make clean
make all # ~9 hrs 25 minutes
\end{lstlisting}

\subsection{Evaluation and expected results}
The artifact scripts redirect the raw output numbers and logs
to the \texttt{output/} folder, while the plotted graphs are saved as PDFs in
the \texttt{graphs/} folder. Summary data for each figure is saved as CSV files in the \texttt{csvs/} directory. Results may have minor runtime variations from those reported in the paper, but general trends should hold.

\subsection{Notes}
\begin{itemize}
  \item We recommend running experiments inside a \texttt{tmux} session, as some benchmarks take several hours.
  \item Model weights for Llama-3.3-70B, Qwen2.5-72B, and Mixtral-8x22B require approximately 500~GB of disk space in total (downloaded and cached by Hugging Face on first run).
  \item Individual experiments can be re-run independently without affecting others. Use \texttt{make clean} to remove all generated outputs before a fresh run.
  \item Plots can be regenerated from existing output data without re-running experiments: \texttt{make gen\_all}.
\end{itemize}

\subsection{Methodology}

Submission, reviewing, and badging methodology:

\begin{itemize}
  \item \url{http://cTuning.org/ae/submission-20190109.html}
  \item \url{http://cTuning.org/ae/reviewing-20190109.html}
  \item \url{https://www.acm.org/publications/policies/artifact-review-badging}
\end{itemize}
\end{document}